%% file: ms.tex
\documentclass[iop]{emulateapj}
\usepackage{lscape}
\usepackage{multirow}
\usepackage{natbib}
\usepackage{placeins}

\newcommand{\spitzer}{\emph{Spitzer}}
\newcommand{\herschel}{\emph{Herschel}}

\newcommand{\degree}{\mbox{$^{\circ}$}}

\newcommand{\kms}{\mbox{km s$^{-1}$}}% km/s
\newcommand{\mjybeam}{\mbox{mJy beam$^{-1}$}}% Jy/beam
\newcommand{\mjysr}{\mbox{MJy sr$^{-1}$}}% Jy/beam

\newcommand{\um}{$\mu$m}
\newcommand{\lsun}{\mbox{L$_\odot$}}% Lsun
\newcommand{\msun}{\mbox{M$_\odot$}}% Msun
\newcommand{\cojone}{$^{12}$CO (1--0)}

\newcommand{\beq}	{\begin{equation}}
\newcommand{\eeq}	{\end{equation}}
\newcommand{\beqa}{\begin{eqnarray}}
\newcommand{\eeqa}{\end{eqnarray}}

\slugcomment{\footnotesize Accepted by ApJ}

\begin{document}
%%%%%%%%%%%%%%%%%% title %%%%%%%%%%%%%%%%%%%%%%%%%%%%%%%%%%%%%%%%
\title {An ALMA Search for Substructure, Fragmentation, and Hidden Protostars in Starless Cores in Chamaeleon~I}

\author{
Michael M.~Dunham\altaffilmark{1,2}, 
Stella S.~R.~Offner\altaffilmark{3}, 
Jaime E.~Pineda\altaffilmark{4}, 
Tyler L.~Bourke\altaffilmark{5}, 
John J.~Tobin\altaffilmark{6},
H\'ector G.~Arce\altaffilmark{7}, 
Xuepeng Chen\altaffilmark{8}, 
James Di Francesco\altaffilmark{9}, 
Doug Johnstone\altaffilmark{9,10},  
Katherine I.~Lee\altaffilmark{1},
Philip C.~Myers\altaffilmark{1},
Daniel Price\altaffilmark{11}, 
Sarah I.~Sadavoy\altaffilmark{12}, 
\& 
Scott Schnee\altaffilmark{13}
}

\altaffiltext{1}{Harvard-Smithsonian Center for Astrophysics, 60 Garden Street, MS 78, Cambridge, MA 02138, USA}

\altaffiltext{2}{mdunham@cfa.harvard.edu}

\altaffiltext{3}{Department of Astronomy, University of Massachusetts, Amherst, MA 01002, USA}

\altaffiltext{4}{Max-Planck-Institut f\"{u}r extraterrestrische Physik, Giessenbachstrasse 1, 85748, Garching, Germany}

\altaffiltext{5}{SKA Organization, Jodrell Bank Observatory, Lower Withington, Macclesfield, Cheshire SK11 9DL, UK}

\altaffiltext{6}{Leiden Observatory, Leiden University, P.O. Box 9513, 2300-RA Leiden, The Netherlands}

\altaffiltext{7}{Department of Astronomy, Yale University, P.O. Box 208101, New Haven, CT 06520, USA}

\altaffiltext{8}{Purple Mountain Observatory, Chinese Academy of Sciences, 2 West Beijing Road, Nanjing 210008, China}

\altaffiltext{9}{National Research Council of Canada, Herzberg Astronomy \& Astrophysics Programs, 5071 West Saanich Road, Victoria, BC, V9E 2E7, Canada}

\altaffiltext{10}{Department of Physics and Astronomy, University of Victoria, Victoria, BC V8P 1A1, Canada}

\altaffiltext{11}{Monash Centre for Astrophysics (MoCA) and School of Mathematical Sciences, Monash University, VIC 3800, Australia}

\altaffiltext{12}{Max-Planck-Institut f\"{u}r Astronomie (MPIA), K\"{o}nigstuhl 17, 69117 Heidelberg, Germany}

\altaffiltext{13}{1National Radio Astronomy Observatory, 520 Edgemont Road, Charlottesville, VA 22903, USA}

\begin{abstract}
We present an Atacama Large Millimeter/submillimeter Array (ALMA) 106 GHz 
(Band 3) continuum survey of the complete population of dense cores in the 
Chamaeleon~I molecular cloud.  We detect a total of 24 continuum 
sources in 19 different target fields.  All previously known Class 0 and Class 
I protostars in Chamaeleon~I are detected, whereas all of the 56 starless cores 
in our sample are undetected.  
We show that the \spitzer+\herschel\ census of protostars in 
Chamaeleon~I is complete, with the rate at which protostellar cores have been 
misclassified as starless cores calculated as $<$1/56, or $<$ 2\%.  
We use synthetic observations to show that starless cores collapsing following 
the turbulent fragmentation scenario are detectable by our ALMA observations 
when their central densities exceed $\sim$10$^8$~cm$^{-3}$, with the exact 
density dependent on the viewing geometry.  Bonnor-Ebert spheres, on the 
other hand, remain undetected to central densities at least as high as 
$10^{10}$~cm$^{-3}$.  Our starless core non-detections are used to infer that 
either the star formation rate is declining in Chamaeleon~I and most of the 
starless cores are not collapsing, matching the findings of previous studies, 
or that the evolution of starless cores are more accurately described by 
models that develop less substructure than predicted by the turbulent 
fragmentation scenario, such as Bonnor-Ebert spheres.  We outline future 
work necessary to distinguish between these two possibilities.
\end{abstract}


%%%%%%%%%%%%%%%%%%%%%%%%%%%%%%%%%%%%%%%%%%%%%%%%%%%%%%%%%%%%%

\section{Introduction}\label{sec_intro}
Dense molecular cloud cores are the eventual birthplaces of stars 
\citep[e.g.,][]{beichman1986:iras,difrancesco2007:ppv}.  Sensitive 
(sub)millimeter bolometers with large instantaneous fields-of-view have 
enabled rapid mapping of entire molecular clouds, and as a result dense cores 
are often identified via their dust continuum emission 
\citep[e.g.,][]{motte1998:oph,shirley2000:scuba,johnstone2000:scuba,johnstone2001:scuba,enoch2006:bolocam,enoch2007:serpens,sadavoy2010:cores,pattle2015:scuba2,salji2015:scuba2,konyves2015:herschel,kirk2016:scuba2}.  
After identification, they are typically classified as 
protostellar or starless based on the presence or absence, respectively, 
of a young stellar object (YSO) detected in the infrared 
\citep[e.g.,][]{benson1984:cores,beichman1986:iras,myers1987:iras,jorgensen2008:scubaspitzer,enoch2009:protostars} 
or a molecular outflow 
\citep[e.g.,][]{hatchell2009:outflows,chen2010:fhsc}.  
Some authors identify a subset of starless cores as ``prestellar cores'' if 
evidence exists that they are gravitationally bound or unstable \citep[see][for reviews of this subject]{difrancesco2007:ppv,wardthompson2007:ppv}.  
In this work, we use the term ``starless core'' to refer to any core not found 
to be associated with a protostar, where protostar refers to a YSO 
still embedded in and accreting from its parent dense core.  
We avoid the term ``prestellar'' altogether since we lack the data to 
evaluate whether each core is bound or unstable.

While existing (sub)millimeter 
observations have been extremely successful at finding 
cores, these relatively low angular resolution single-dish continuum surveys 
have been unable to determine the detailed physical structure of starless 
cores, especially the substructures indicative of fragmentation that will 
eventually form multiple protostellar systems.  Additionally, infrared 
observations with the {\it Spitzer Space Telescope} 
\citep[{\it Spitzer};][]{werner2004:spitzer} and 
{\it Herschel Space Observatory} 
\citep[{\it Herschel};][]{pilbratt2010:herschel} 
can miss the most deeply embedded and lowest luminosity protostars, leading 
to incorrect classification of some protostellar cores as starless.

\subsection{Fragmentation and Substructure in Starless Cores}

Approximately half of all protostars are found in multiple systems 
\citep{looney2000:multiplicity,maury2010:pdbi,chen2013:multiplicity,tobin2016:vandam}.  Additionally, both starless and protostellar cores are 
often found to have irregular morphologies on $\sim$1000 AU scales based on 
mid-infrared extinction studies 
\citep[e.g.,][]{stutz2009:starless,tobin2010:protostars}, suggesting conditions 
favorable for fragmentation.  Indeed, simulations including the effects of 
radiative feedback predict that ``turbulent fragmentation'' -- the process 
by which turbulent fluctuations in a dense core become Jeans unstable and 
collapse faster than the background core 
\citep[e.g.,][]{fisher2004:turbfrag,goodwin2004:turbfrag} -- is the dominant 
mechanism responsible for forming multiple systems, and that 
the fragmentation begins in the starless core stage 
\citep{offner2010:turbfrag}.  

To date, only a few, isolated cases of substructure and 
fragmentation in starless cores have been identified \citep[e.g.,][]{kirk2009:l483,chen2010:cra,nakamura2012:starless,bourke2012:starless,lee2013:starless,friesen2014:starless,pineda2011:b5,pineda2015:b5}.  
On the other hand, the largest interferometric survey of starless cores 
presented to date, a CARMA (Combined Array for Research in Millimeter-wave 
Astronomy) 3 mm continuum survey of 12 cores in Perseus, found no 
evidence for substructure and fragmentation 
\citep{schnee2010:starless,schnee2012:starless}.  However, this CARMA survey 
was only sensitive to compact structures with masses above 
$\sim 0.1$ \msun, and synthetic observations of the \citet{offner2010:turbfrag} 
turbulent fragmentation simulations have shown that CARMA likely lacked the 
sensitivity necessary to detect substructure and fragmentation 
\citep{offner2012:simalma,mairs2014:simalma}.  These same synthetic 
observations have shown that the Atacama Large Millimeter/submillimeter Array 
(ALMA) is capable of detecting substructure and fragmentation in starless 
cores, particularly once it enters full science operations.

\subsection{Accurate Classification of Cores}

The first starless core observed by \spitzer\ in the Cores to Disks Legacy 
Survey \citep[c2d;][]{evans2003:c2d,evans2009:c2d}, L1014, turned out to harbor 
a low luminosity, embedded protostar \citep{young2004:l1014}.  In a full 
search of the c2d dataset, \citet{dunham2008:lowlum} found that $\sim$20\% 
of the ``starless cores'' in c2d contained similarly low-luminosity protostars 
and were thus misclassified.  They further postulated the existence of an even 
lower luminosity population of protostars below the sensitivity limit of 
\spitzer, based on the observation that they continued to detect protostars 
all the way down to their sensitivity limit of 0.01 \lsun\ at $d=230$ pc.  
Such extremely faint protostars have started to be found in recent years, 
most through interferometer detections of outflows and/or compact 
continuum emission from cores classified 
as starless in \spitzer\ observations \citep[e.g.,][]{enoch2010:fhsc,chen2010:fhsc,chen2012:fhsc,pineda2011:fhsc,dunham2011:fhsc}.
All of these newly discovered objects have been suggested as candidate 
first hydrostatic cores, a short-lived stage intermediate between the starless 
and protostellar stages beginning when the first central, hydrostatic object 
forms and lasting until the central temperature reaches $\sim$2000 K and 
H$_2$ dissociates \citep{larson1969:fhsc}.  None of these candidates, however, 
have been reliably 
confirmed as first hydrostatic cores due to a lack of clear theoretical 
predictions for distinguishing between first hydrostatic cores and very young 
protostars \citep[see][for a recent review]{dunham2014:ppvi}.  For simplicity 
we will refer to them here as ``newly identified'' or ``new'' protostars while 
acknowledging that some may in fact be first hydrostatic cores.

The detection rate of these new protstars has been slow due to the 
sensitivity limitations of facilities like CARMA and the Submillimeter Array 
(SMA), which typically require one or more full tracks of observations to 
achieve sufficient sensitivity.  However, even with the limited observations 
available to date, at least 5--20\% of cores in the Perseus molecular cloud 
that are ``starless'' even to \spitzer\ sensitivities have been found to harbor 
deeply embedded, low-luminosity protostars \citep{schnee2010:starless}.  
While \herschel\ 
has also found additional protostars missed by \spitzer\ 
\citep[e.g.,][]{stutz2013:pbrs,sadavoy2014:perseus}, most of the new protostars 
identified through interferometric observations of supposedly starless 
cores remain undetected even by \herschel\ \citep{pezzuto2012:fhsc}.  
Thus sensitive interferometric observations of large numbers of starless 
cores are needed to obtain a full census of protostars in nearby clouds.

Obtaining an accurate census of the numbers of cores in the various 
stages or classes of star formation, and thus an accurate measure of the 
relative and absolute lifetimes of each stage or class, provide important 
constraints on theories 
of star formation. For instance, theoretical predictions of the lifetime of 
starless cores are uncertain by greater than one order of magnitude, with 
estimates ranging from 0.1 Myr to greater than 1 Myr 
\citep[see][for a recent review]{mckee2007:review}.  Observations from 
the \spitzer\ c2d survey suggest a starless core lifetime of 
$\sim$0.5 Myr \citep{enoch2008:starless}, intermediate between the extremes 
predicted by theory, though this estimate is biased by c2d's lack of 
sensitivity to the faintest protostars.  With an accurate census of starless 
cores, their lifetimes can be determined, providing direct constraints on 
theories of star formation.

Additionally, accretion theories can be tested against the observed 
distribution of protostellar luminosities 
\citep[e.g.,][see Dunham et al.~2014 for a recent review]{dunham2010:evolmodels,dunham2012:evolmodels,offner2011:luminosities,padoan2014:luminosities}.  
However, current observed protostellar luminosity distributions are based on 
\spitzer\ observations \citep{dunham2008:lowlum,dunham2013:luminosities,evans2009:c2d,kryukova2012:luminosities} and are
thus incomplete at the lowest luminosities, exactly those luminosities that 
most strongly constrain the underlying accretion theories.  
At present, the low end of the protostellar luminosity distribution is 
relatively undersampled by observations, leaving theoretical predictions on
the lower limits of accretion luminosity and the number of such faint objects 
rather unconstrained.

\subsection{An ALMA Survey of Chamaeleon~I}

\begin{figure*}
\epsscale{0.9}
\plotone{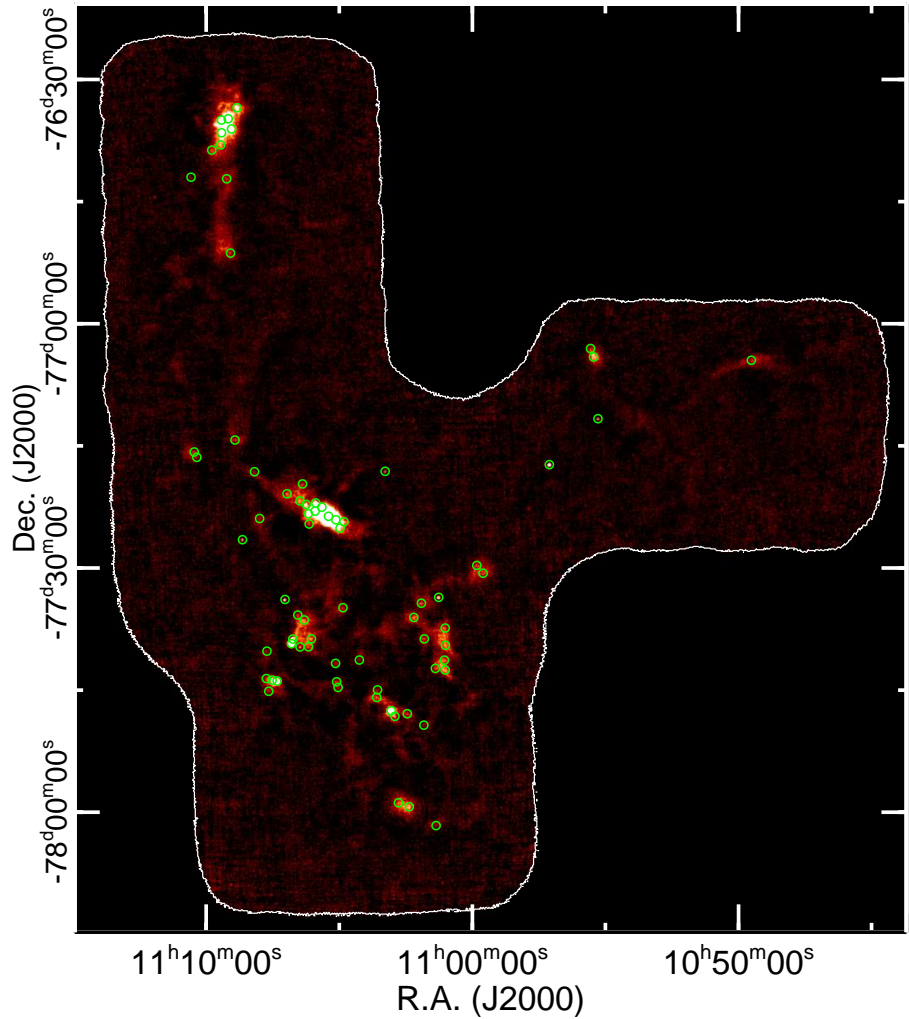}
\vspace{-0.35in}
\caption{\label{fig_cloud}APEX LABOCA 870 \um\ image of the Chamaeleon~I cloud 
from \citet{belloche2011:chami}.  The image is displayed on a linear scale 
ranging from $-$0.1 (black) to 0.15 (white) Jy~beam$^{-1}$.  The small green 
circles show the 73 ALMA pointings presented in this paper, with the diameters 
of the circles equal to the FWHM of the primary beam.}
\end{figure*}

Sensitive interferometric surveys of large numbers of starless cores 
are required to make progress on improving detection statistics.  
Such surveys will 
constrain the level of substructure and fragmentation that develops in the 
starless core stage, and they will accurately determine the true number of 
starless cores, thus constraining both their lifetimes and the accretion 
luminosities and histories of the youngest protostars.  With the advent of 
ALMA, such surveys are now possible.  In this paper, we present the results 
of a Band 3 (106 GHz; 2.8 mm) ALMA Cycle 1 survey of all known dense cores in 
the Chamaeleon~I molecular cloud, which is located at a distance of 150 pc 
\citep[see discussion of distance in][]{belloche2011:chami}.  
We target all 73 sources detected by \citet{belloche2011:chami} in a 
single-dish 870 \um\ continuum survey with the Large Apex Bolometer Camera 
(LABOCA) at the Atacama Pathfinder Experiment (APEX).  With ALMA's exquisite 
sensitivity, our survey targets six times more cores to ten times better mass 
sensitivity than the \citet{schnee2010:starless} CARMA survey of Perseus cores 
in a fraction of the total observing time, and lays the groundwork for 
future surveys targeting similar populations in other star-forming clouds.

The organization of our paper is as follows:  In \S \ref{sec_observations}, 
we describe the observations and data reduction.  We present our basic 
results in \S \ref{sec_results}, including a summary of the properties of 
the detected sources and an overview of the evolutionary status of each core, 
incorporating information from our ALMA observations.  We discuss the 
implications of our results in \S \ref{sec_implications_census} and 
\S \ref{sec_implications_starless}, focusing on developing an accurate census 
of starless and protostellar cores in \S \ref{sec_implications_census} 
and constraints on substructure and fragmentation in starless cores in 
\S \ref{sec_implications_starless}.  Finally, we summarize our results 
in \S \ref{sec_summary}.

\section{Observations}\label{sec_observations}

We obtained ALMA observations of every source in Chamaleon I detected 
in the single-dish 870 \um\ LABOCA survey by \citet{belloche2011:chami} 
except for those listed as likely artifacts (1 source), residuals 
from bright sources (7 sources), or detections tentatively
 associated with young stellar objects (3 sources).  The first two categories 
were omitted to avoid unreliable sources, whereas the third category was 
omitted due to constraints on the maximum number of pointings allowed in a 
single observing program.  Thus, we observed 73 sources 
from the initial list of 84 objects identified by \citet{belloche2011:chami}.  
Figure \ref{fig_cloud} shows the 73 ALMA pointings overlaid on the 
APEX LABOCA 870 \um\ image of Chamaeleon~I presented by 
\citet{belloche2011:chami}.

We observed the 73 pointings using the ALMA Band 3 
receivers during its Cycle 1 campaign between 2013 November 29 and 2014 
March 08.  Between 25 and 27 antennas were available 
for our observations, with the array configured in a relatively compact 
configuration to provide a resolution of approximately 2\arcsec\ FWHM 
(300 AU at the distance to Chamaeleon~I).  Each target 
was observed in a single pointing with approximately 1 minute of on-source 
integration time.  
Three out of the four available spectral windows were 
configured to measure the continuum at 101, 103, and 114 GHz, each with a 
bandwidth of 2 GHz, for a total continuum bandwidth of 6 GHz (2.8 mm) at a 
central frequency of 106 GHz.  These continuum observations achieved a 
1$\sigma$ rms noise of approximately 0.1 mJy\,beam$^{-1}$. 
The remaining window was configured to observe the 
\cojone\ line at 115 GHz.  Only the continuum data are presented here.  The 
maximum angular scale that can be recovered in these ALMA data is 
$\sim$25\arcsec.  

The calibration and imaging were done using the Common Astronomy Software 
Applications (CASA) package\footnote{Available at http://casa.nrao.edu} 
following the standard routines described in 
\citet{petry2014:alma} and \citet{schnee2014:alma}.  Additional refinement
of the brightest continuum detections (those brighter than $\sim$3~mJy) was 
carried out through two rounds of self-calibration, the first phase-only and 
the second with amplitude and phase corrections.  Overlapping fields were 
mosaiced together.

Table \ref{tab_observations} lists, for each source, 
an identifier using the detection number from \citet{belloche2011:chami}, the 
phase center of our ALMA observations (taken directly from the core positions 
given by Belloche et al.), the FWHM size and position angle of 
our ALMA continuum synthesized beam, the 1$\sigma$ rms of our ALMA 
continuum data and corresponding 1$\sigma$ mass sensitivity, and a flag 
noting the evolutionary status of each source as determined by us (see 
discussion in \S \ref{sec_results_census} for more details regarding 
source evolutionary status).  The 1$\sigma$ mass sensitivities are 
calculated assuming emission from optically thin, isothermal dust (see 
Equation \ref{eq_dustmass} and accompanying text below for more details on 
this calculation).

\section{Results}\label{sec_results}

\subsection{Continuum Detections: Observed and Physical Properties}

With a pixel size of 0.25\arcsec, there are approximately 181,000 pixels 
over the primary beam (approximately 60\arcsec).  With such a large 
number of pixels, Gaussian statistics predict over 500 pixels above 
3$\sigma$ simply due to random noise, but $<$1 above 5$\sigma$.  To avoid 
spurious detections, we thus adopt 5$\sigma$ as the minimum significance 
required for detection in this paper.

We detect a total of 24 106 GHz continuum sources to 5$\sigma$ or greater 
significance in 19 different target 
fields, leaving 54 target fields with no detections.
%\footnote{One additional 
%5$\sigma$ detection is tentatively seen in the CHAI-39 pointing, but as it 
%is outside of the full-width at 10\% power point of the primary beam it 
%is not a sufficiently robust detection to report here.}  
Table \ref{tab_continuum} lists the name of each detected source (taken to be 
the same as the target field name) along with the results from an elliptical 
Gaussian fit to each source using the {\sc imfit} task in CASA, including 
the source position, major and minor axes and position angle after 
deconvolution with the beam, peak flux density, and integrated flux density.  
All measurements were made on images corrected for primary beam attenuation.  
Fields with multiple sources detected append `A', `B', etc. to each source in 
alphabetical order with increasing Right Ascension.  Sources found to be 
unresolved in one or both dimensions and thus unable to be deconvolved with the 
beam are noted.  

Table \ref{tab_continuum_physical} lists the physical properties of each 
detected source, including the effective radius, total mass, and mean total gas 
number density.  
The effective radius, $r_{eff}$, is calculated as the geometric mean of the 
semimajor and semiminor axes of the source, if beam deconvolution is possible, 
or the beam if not (in which case it is expressed as an upper limit).  
The mass is calculated as

\begin{figure*}[hbt]
\epsscale{1.13}
\plotone{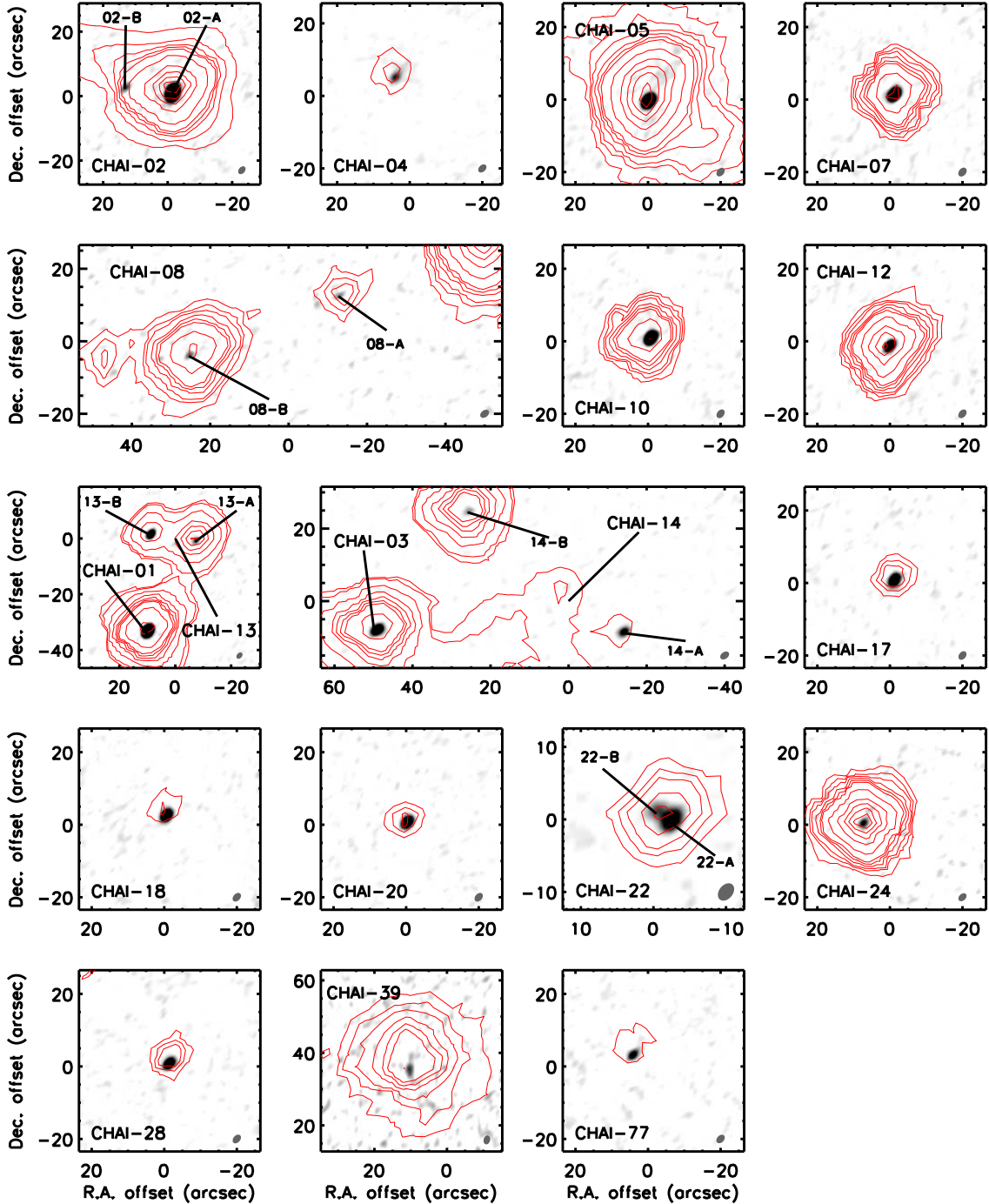}
\vspace{-0.35in}
\caption{\label{fig_herschel}Images of each source detected by our ALMA 
observations.  The images are uncorrected for primary beam attenuation for 
display purposes.  In each panel, the grayscale shows the ALMA 106 GHz 
continuum emission and the red contours show \herschel\ 70 \um\ emission from 
the \herschel\ Gould Belt survey, which were observed with a beam 
full-width at half-maximum (FWHM) of 8.4$''$ \citep{konyves2015:herschel}. 
The gray ellipse at the lower right of each panel shows the 
ALMA synthesized beam.  Each panel is 
labeled with the name of the target field in which the source is detected, 
with the coordinate offsets displayed relative to the core centers (which are 
also the phase centers of our observations; see Table \ref{tab_observations}).  
Panels covering multiple cores 
have the central positions of each core marked, and panels with multiple 
sources detected have each source labeled.  The grayscale is plotted on a 
linear scale ranging from 0.1 (white) to 2 (black) \mjybeam\ for each 
panel except CHAI-08, CHAI-24, and CHAI-39, where the maximum is 
instead 1 \mjybeam.  In all panels except for five, the \herschel\ 70 \um\ 
contours are plotted at 10, 15, 20, 30, 40, 50, 100, 200, 300, 400, 500, 1000, 
3000, 5000, 7000, and 9000 \mjysr.  For the following five panels, the same 
pattern is kept except with different minimum contour levels:  CHAI-02 (100 
\mjysr), CHAI-13 and CHAI-14 (40 \mjysr), and CHAI-18 and CHAI-77 (5 \mjysr).}
\end{figure*}

\begin{figure*}[hbtp]
\epsscale{1.25}
\plotone{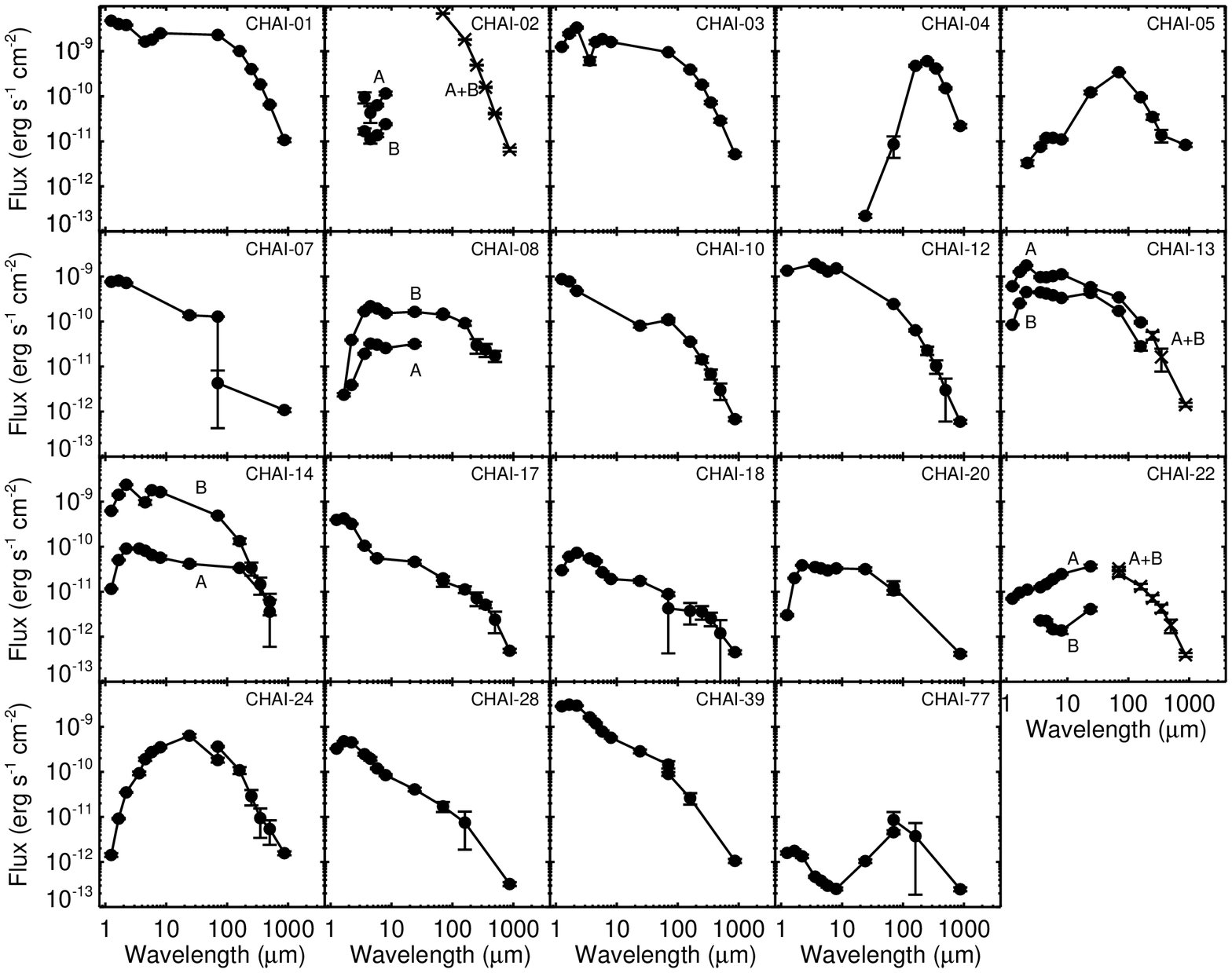}
\caption{\label{fig_seds}Spectral energy distributions (SEDs) for each of the 
24 sources detected by our ALMA 106 GHz continuum observations, based on 
associations with 2MASS, \spitzer, \herschel, and LABOCA sources.  The 
associated 2MASS and \spitzer\ sources are taken from the YSO catalogs 
tabulated by the \spitzer\ Gould Belt Legacy Survey \citep{dunham2015:gb}, 
and the associated \herschel\ sources are taken from the \herschel\ Gould Belt 
Survey as tabulated by \citet{winston2012:chami}.  Each panel is labeled with 
the target field in which the source is detected.  Fields with multiple 
detected sources have each SED separately plotted as filled circles and 
labeled when resolved, and the combined SED plotted with an x (and labeled) 
when unresolved.  We do not include our ALMA continuum measurements since they 
may suffer from spatial filtering.}
\end{figure*}

\begin{equation}\label{eq_dustmass}
M = 100 \frac{d^2 S_{\nu}}{B_{\nu}(T_D) \kappa_{\nu}} \quad ,
\end{equation}
where $S_{\nu}$ is the integrated flux density, $B_{\nu}(T_D)$ is the Planck 
function at the isothermal dust temperature $T_D$, $\kappa_{\nu}$ is the dust 
opacity, $d = 150$ pc, and the factor of 100 is the assumed gas-to-dust 
ratio.  We adopt the dust opacities of \citet{ossenkopf1994:oh5} appropriate 
for thin ice mantles after $10^5$ yr of coagulation at a gas density of $10^6$ 
cm$^{-3}$ (OH5 dust), giving $\kappa_{\nu} = 0.23$ cm$^2$ g$^{-1}$ at 106 GHz.  
With the integrated flux densities listed in Table \ref{tab_continuum} and an 
assumed $T_D = 10$ K, Equation \ref{eq_dustmass} gives total masses ranging 
from 0.012 \msun\ to 1.7 \msun.  Note that the uncertainties listed in Table 
\ref{tab_continuum_physical} only include the statistical uncertainties in the 
integrated flux densities.  The true uncertainties are likely dominated by the 
dust temperature and opacity assumptions.  As 3 mm is strongly in the 
Rayleigh-Jeans limit, a factor of two uncertainty in dust temperature 
directly translates into a factor of two uncertainty in mass.  For the opacity, 
typical uncertainties of factors of $2-4$ are quoted based on the specific 
dust model \citep[e.g.,][]{shirley2005:l1498,shirley2011:opacity}.  However, 
the opacities can in fact be larger by an order of magnitude or more if dust 
grains have grown to millimeter sizes or larger, as such large grains will 
flatten the dust opacity power-law index \citep[e.g.,][]{ricci2010:disks,tobin2013:l1527,testi2014:ppvi,schnee2014:beta}.  
Shallower dust opacity power-law indices caused by larger grains would 
systematically increase the opacities compared to the assumed OH5 value, 
decreasing the calculated masses.  While grains are not generally expected 
to grow to such large sizes in starless cores, \citet{kwon2009:beta} and 
\citet{schnee2014:beta} did recently find evidence for large grains in both 
large-scale filaments and dense cores, although the Schnee et al.~results 
have currently been questioned by \citet{sadavoy2016:beta}.  If large dust 
grains in dense starless cores are confirmed by future studies, the opacities 
assumed here will require upward revision.

The mean total gas number density, $n$, is calculated assuming spherical 
symmetry as

\begin{equation}\label{eq_numberdens}
n = \frac{3}{4 \pi \mu m_H} \frac{M}{r^3_{eff}}
\end{equation}
where $M$ is the mass, $r_{eff}$ is the effective radius, $m_{H}$ is the 
hydrogen mass, and $\mu$ is the mean molecular weight per free particle.\footnote{If we instead used $\mu_{\rm H_2}$, the mean molecular weight per hydrogen molecule ($\mu_{\rm H_2} = 2.8$ for gas that is 71\% by mass hydrogen, 27\% helium, and 2\% metals; \citealt{kauffmann2008:mambo}), Equation \ref{eq_numberdens} would give the H$_2$ mean number density rather than the total gas mean number density.}
With $M$ and $r_{eff}$ as calculated above and $\mu = 2.37$ for gas that is 
71\% by mass hydrogen, 27\% helium, and 2\% metals 
\citep{kauffmann2008:mambo}, we calculate and tabulate number densities in 
the last column of Table \ref{tab_continuum_physical}.  As with the mass, 
only the statistical uncertainties are listed in Table 
\ref{tab_continuum_physical}; the true uncertainties are dominated by 
systematic uncertainties in the assumed distance, source geometries, dust 
temperatures, and opacities, as discussed above.

Figure \ref{fig_herschel} shows images of each detected source overlaid with 
contours of the \herschel\ 70 \um\ emission from the Gould Belt 
survey\footnote{Available at http://gouldbelt-herschel.cea.fr} 
\citep{andre2010:gb,winston2012:chami}.  Figure \ref{fig_seds} shows 
spectral energy distributions (SEDs) for each source constructed from 
2MASS$+$\spitzer, \herschel, and LABOCA photometry tabulated by 
\citet{dunham2013:luminosities,dunham2015:gb}, \citet{winston2012:chami}, and 
\citet{belloche2011:chami}, respectively.  In the latter case the total 
flux densities rather than peak intensities are plotted.  
We do not include our ALMA 106 
GHz continuum measurements in Figure \ref{fig_seds} since they likely suffer 
from unknown levels of spatial filtering compared to the single-dish data.

\subsection{Associations with Detections at Other Wavelengths}\label{sec_results_associations}

In order to investigate the nature of the continuum detections, we now 
consider their associations with sources detected at other wavelengths.  

\subsubsection{Association with LABOCA Sources}

Column 2 of Table \ref{tab_associations} lists the projected separation 
between each ALMA detection and the associated 870 \um\ LABOCA source 
targeted by our ALMA observations.  We consider our sources to be physically 
associated with the corresponding LABOCA sources if they are located 
within a projected separation of 10.6\arcsec\ \citep[the half-power radius of 
the 21.2\arcsec\ FWHM beam of the LABOCA observations;][]{belloche2011:chami}.  
Associations following this definition are tabulated in column 3 of Table 
\ref{tab_associations}, showing that 18 of the 24 ALMA 106 GHz continuum 
detections are associated with LABOCA sources.  Despite the ALMA 
detections of sources in the fields targeting CHAI-08, CHAI-14, and CHAI-39, 
these ALMA detections are located far enough away from the 
LABOCA sources themselves that we do 
not consider them to be physically associated with these sources.

\subsubsection{Association with \herschel\ YSOs}

Columns 4--6 of Table \ref{tab_associations} list the YSO from 
\citet{winston2012:chami} associated with each detected source, along with the 
projected separation and evolutionary class 
(the latter taken directly from their Table 1).  
Winston et al.~start with a list of 397 sources extracted from the \herschel\ 
70--500 \um\ images of Chamaeleon~I obtained as part of the \herschel\ Gould 
Belt Legacy Survey \citep{andre2010:gb}.  They associated 49 of these 
\herschel\ detections with YSOs previously identified with \spitzer\ by 
\citet{luhman2008:chami} and list \herschel\ flux densities for those sources.  
All but two of our 106 GHz ALMA continuum detections are associated with these 
YSOs.  One is a Class 0 object, five are Class I objects, four are 
flat-spectrum objects, 11 are Class II objects, 
and one is a transition disk 
\citep[see][for details]{winston2012:chami}.  
We thus detect all of the Class 0 and I YSOs listed by 
\citet{winston2012:chami}, and four out of the six flat-spectrum 
objects.  
The remaining two flat-spectrum objects are not covered by our 73 
ALMA pointings.

The two ALMA detections not associated with YSOs from 
\citet{winston2012:chami} are CHAI-02-B and CHAI-08-A.  CHAI-02-B is located 
13.5\arcsec\ east of CHAI-02-A.  Visual inspection of the \spitzer\ images 
provided by the \spitzer\ Gould Belt survey \citep{dunham2015:gb}
reveals a mid-infrared point source detected at 3.6-8 \um, and indeed it is 
listed as a flat-spectrum YSO by \citet{luhman2008:chami}.  It is not resolved 
into a separate source at \herschel\ wavelengths, explaining its absence 
from the \citet{winston2012:chami} catalog.  CHAI-08-A is associated with a 
flat-spectrum YSO detected by \spitzer\ \citep{luhman2008:chami}, but it is 
not listed in the \citet{winston2012:chami} catalog.

\subsubsection{Association with \spitzer\ Protostars}

Columns 7--9 of Table \ref{tab_associations} list the associated protostar 
from the \citet{dunham2013:luminosities} compilation of protostars detected 
by \spitzer\ in the Gould Belt clouds, including Chamaeleon~I.  This protostar 
catalog is a subset of the full Gould Belt YSO catalog presented by 
\citet{dunham2015:gb}.  Only 9 out 
of the 24 ALMA detections are associated with \spitzer-identified protostars.  
\citet{dunham2013:luminosities} emphasized reliability over completeness 
by requiring detections in all five \spitzer\ bands between 3.6--24 \um\ 
and association with dense cores as traced by complementary (sub)millimeter 
continuum surveys.  Of the 15 ALMA detections that are not associated with 
protostars listed by \citet{dunham2013:luminosities}, four are saturated at 
one or more wavelengths, three are near \spitzer\ map edges and thus not 
covered at one or more wavelengths, one is not detected at one or more 
wavelengths, three are not resolved at one or more wavelengths, and four are 
detected at all \spitzer\ wavelengths but not projected onto the positions of 
dense cores.  We note here that the only Class 0 protostar in Chamaeleon~I 
(CHAI-04, also known as Cha-MMS1) is not in the Dunham et al.~catalog as it is 
not detected at all \spitzer\ wavelengths.  Indeed, this object has been 
identified as a candidate first hydrostatic core \citep{larson1969:fhsc,belloche2006:chammms1,tsitali2013:chammms1,vaisala2014:chammms1}.  
Furthermore, two of the five Class I 
protostars in Chamaeleon~I are also missing from the Dunham et al.~catalog: 
CHAI-02-A due to saturation and CHAI-22-B due to not being 
resolved at all wavelengths.  While \citet{dunham2013:luminosities} did 
acknowledge that their sample focuses on reliability at the expense of 
completeness, we emphasize here that the list of protostars in nearby clouds 
must continue to be updated as new information becomes available.

\subsection{Evolutionary Status of the 73 Targets}\label{sec_results_census}

The last column of Table \ref{tab_observations} notes the evolutionary status 
of each of the 73 targets, determined as described below.  

As noted above, 19 target fields feature continuum detections.  Of those, 
the one associated with a Class 0 source (CHAI-04) and the four associated 
with Class I sources (CHAI-02, CHAI-05, CHAI-22, and CHAI-24) are clearly 
protostars, thus these objects are marked as protostellar cores.  Eleven 
additional detections are associated (at least in projection) with 
LABOCA sources: 
CHAI-01, 03, 07, 10, 12, 13, 17, 18, 20, 28, and 77.  However, 
as evident from Figure \ref{fig_seds} and Column 6 of Table 
\ref{tab_associations}, all of these objects exhibit flat-spectrum or Class II 
SEDs indicative of more evolved objects near or beyond the end of the 
embedded, protostellar stage of evolution.  
One possible explanation for these sources being associated with LABOCA 
sources is that they are protostars viewed nearly 
face-on down outflow cavities, as such objects could exhibit Class II SEDs 
\citep[e.g.,][]{whitney2003:geometry1,dunham2012:evolmodels}.  However, such 
objects tend to feature double-peaked SEDs 
\citep[e.g.,][]{whitney2003:geometry1} 
not seen in Figure \ref{fig_seds}.  
Another, more likely possibility is that the associated LABOCA detections are 
not tracing dense cores. The \citet{belloche2011:chami} 
LABOCA survey has a mass sensitivity more than six times 
deeper than surveys such as the Bolocam 1.1 mm surveys of 
nearby clouds \citep{enoch2006:bolocam,young2006:ophiuchus,enoch2007:serpens}.  
It is also capable of recovering extended emission on scales up to a factor of 
2--5 larger than other, previous surveys \citep[see Appendix A of][for a detailed discussion of spatial recovery]{belloche2011:chami}.  Thus, the LABOCA 
survey is more likely to detect both extended dust emission and dust in disks 
that are no longer embedded within dense cores, and indeed, for these 
reasons, \citet{belloche2011:chami} never claimed all of their detections were 
tracing dense cores.  While it is possible that some 
of these 11 LABOCA sources are evolved protostars near the 
end of the embedded stage 
\citep[particularly the flat-spectrum sources; e.g.,][]{heiderman2015:misfits}, 
we classify these objects as disks rather than starless or protostellar dense 
cores. Finally, an additional three target fields contain detections that are 
too distant from the LABOCA sources themselves (which are 
located at the zero offset coordinates in Figure \ref{fig_herschel}) to be 
considered associated:  CHAI-08, 14, and 39.  These detections are associated 
with other YSOs located within the ALMA primary beam, and the 
LABOCA sources themselves are considered starless cores.

With one exception, all other LABOCA sources not associated with any 
ALMA 106 GHz continuum detections are considered starless cores, 
leaving us with 56 starless cores, none of which are detected by our ALMA 
continuum observations.  The one exception is CHAI-14: 
the two ALMA detections in this field are unassociated with the core, and the 
core itself is undetected.  However, the LABOCA source is located at a 
projected distance of 3.6\arcsec\ from a Class II YSO from the 
\citet{winston2012:chami} catalog.  As this is well within the half-power 
point of the LABOCA beam, we classify CHAI-14 as a disk rather than a dense 
core, with our ALMA non-detection indicating a low-mass disk with a 3$\sigma$ 
mass limit of 4.2 $\times$ 10$^{-3}$ \msun\ based on the mass sensitivity 
tabulated in Table \ref{tab_observations}.  

\section{A Complete Census of Star Formation in Chamaeleon~I}\label{sec_implications_census}

As noted in \S \ref{sec_intro}, a number of very low luminosity protostars 
have been identified through interferometric detections of outflows and/or 
compact continuum emission from dense cores originally classified as starless 
\citep[e.g.,][]{enoch2010:fhsc,chen2010:fhsc,chen2012:fhsc,schnee2010:starless,schnee2012:starless,pineda2011:fhsc,dunham2011:fhsc}.  
For the most part, these newly identified protostars have been found 
serendipitously, raising the question of exactly how many ``starless'' cores 
are truly starless.  In order to answer this question in Chamaeleon~I 
we must first determine whether or not 
similar protostars would be detected in our observations.  To address 
this, we take four such objects in the Perseus molecular cloud that have been 
observed with interferometers at millimeter wavelengths, with 
synthesized beams ranging from approximately 1$''$ to approximately 6$''$:  
L1448-IRS2E \citep{chen2010:fhsc}, Per-Bolo 45 \citep{schnee2010:starless}; 
Per-Bolo 58 \citep{enoch2010:fhsc,dunham2011:fhsc}, and 
L1451-mm \citep{pineda2011:fhsc}.  We then scale the observed peak intensities 
from 230 pc 
\citep[the distance to Perseus,][]{hirota2008:perseus,hirota2011:perseus} to 
150 pc \citep[the distance to Chamaeleon~I; see][and references therein]{belloche2011:chami}, as well as from the observed 
frequency to 106 GHz assuming $I_{\nu} \propto \nu^{2+\beta}$, where $\beta$ 
is the index of the dust opacity power-law through the far-infrared and 
(sub)millimeter wavelength range.  
The above expression is valid as long as the Rayleigh-Jeans limit 
($h \nu << k T$) is satisfied, which, for $\nu = 106$~GHz, is satisfied for 
$T >> 5.1$K.  
The sources are assumed to have the same dust temperatures in 
Chamaeleon~I as in Perseus since we are simulating observing, at a different 
frequency, these exact sources in a cloud located at a different distance.

The results of scaling the observed peak intensities as described 
above, over the range $0 \leq \beta \leq 2$, are given in Table 
\ref{tab_fhsc} under the assumption that these observed peak intensities 
are dominated by compact, unresolved emission and thus independent of beam 
size.  
After scaling to the distance of Chamaeleon~I, the predicted 106 GHz peak 
intensities range from 0.7 to 17.7 \mjybeam\ depending on the source and the 
assumed value of $\beta$.  As our sensitivities range from 0.08 to 0.14 
\mjybeam, even in the most pessimistic scenario these objects would be 
detected to at least a 5$\sigma$ level, ranging to $>$200$\sigma$ in the most 
optimistic scenario.  Robust detection of first hydrostatic cores is also 
predicted from the comparison with simulations presented below in the 
following section.

Based on these results, we conclude that the \spitzer$+$\herschel\ census of 
protostars in Chamaeleon~I is complete and there are no missing protostars 
or first hydrostatic cores in this cloud.  
The one caveat to this statement is if low-mass, compact cores 
harboring protostars are missing from the \citet{belloche2011:chami} 
catalog due to confusion with larger-scale emission, a possibility we 
cannot rule out here since, as seen in Figure \ref{fig_cloud}, there are 
regions with detected 870~\um\ emission that are not covered by our ALMA 
pointings.  If we assume that the Belloche et al.~catalog is in fact 
complete, with 56 starless cores, we  
find that the rate at which protostellar cores in Chamaeleon~I have been 
misclassified as starless cores is $<$1/56, or $<$ 2\%.

\section{Substructure and Fragmentation in Starless Cores}\label{sec_implications_starless}

As noted in \S \ref{sec_intro}, if fragmentation begins in the starless 
core stage, it should be detectable by ALMA 
\citep{offner2012:simalma,mairs2014:simalma}.  At first 
glance, our non-detections seem to imply that the starless cores in 
Chamaeleon~I are not undergoing a process of turbulent fragmentation like 
that predicted in the simulations of \citet{offner2010:turbfrag}.  
However, the synthetic ALMA observations of the Offner et al.~simulations 
presented by \citet{offner2012:simalma} and \citet{mairs2014:simalma} only 
examined selected times late in the evolution of the starless cores when they 
had already reached relatively high central densities and were close to the 
formation of central, hydrostatic objects.  
In order to fully examine the implications of our non-detections, we must 
determine how far back in time such starless cores are detectable and 
use these results to generate testable predictions for the number of 
detected cores expected in our sample.  To accomplish this task, we revisit 
synthetic ALMA observations of the Offner et al.~simulations, using updated 
simulations that include magnetic fields.

\subsection{Synthetic ALMA Observations of Starless Core Simulations}\label{sec_simulations}

\subsubsection{Description of the Simulations}

We perform two magneto-hydrodynamic (MHD) simulations of isolated, collapsing 
starless cores. The simulations are similar in their properties except for 
initial core masses that differ by one order of magnitude, with the lower 
mass simulation starting with an initial density 2.7 times higher (the higher 
density is required in order for this core to be sufficiently unstable to 
collapse).  
In the first simulation, which we label the C04 simulation, the initial core 
mass is 0.4 \msun.  In the second, which we label the C4 simulation, the 
initial core mass is 4 \msun.  
We use the {\sc orion} Adaptive Mesh Refinement (AMR) code to 
generate self-consistent, time-dependent physical conditions 
\citep{klein99,truelove98,li12}. The simulations include self-gravity and 
determine the gas temperature via a barotropic equation of state (EOS).  The 
EOS models the transition from isothermal to adiabatic as the gas becomes 
optically thick to infrared radiation: the gas pressure is 
$P = \rho c_s^2[1+(\rho/\rho_c)^{\gamma-1}]$, where $c_s$ is the sound speed, 
$\gamma=5/3$ is the ratio of specific heats, and $\rho_c = 2 \times 10^{-13}$ 
g cm$^{-3}$ is the critical density \citep{masunaga98}.  {\sc orion} evolves 
the magnetic field using a constrained transport ideal MHD scheme assuming 
the gas and magnetic field are perfectly coupled \citep{li12}, which is based 
on the finite volume formalism as implemented by \citet{mignone12}. 

The simulations begin with a uniform density spherical core of 10 K gas, 
with initial number densities of $1.6 \times 10^5$ cm$^{-3}$ for the C04 
simulation and $6.0 \times 10^4$ cm$^{-3}$ for the C4 simulation.  The core 
is surrounded by a warm, low density medium with a temperature of 1000 K 
and a density 100 times lower than the initial core density.  The domain 
extent is twice the diameter of the core, and the domain edge has outflow 
boundary conditions. A uniform magnetic field threads the gas in the $z$ 
direction with a strength set such that the mass-to-flux ratio 
is five times the critical value.
At $t=0$, the gas velocities in the core are perturbed with a 
turbulent random field, which has a flat power spectrum over wavenumbers 
$k=1-2$ and is normalized to satisfy the input velocity dispersion. Once set 
in motion, the initial turbulence is allowed to decay and no additional energy 
injection occurs. Table \ref{simprop} summarizes the simulation properties.

\begin{figure*}[hbtp]
\epsscale{0.94}
\plotone{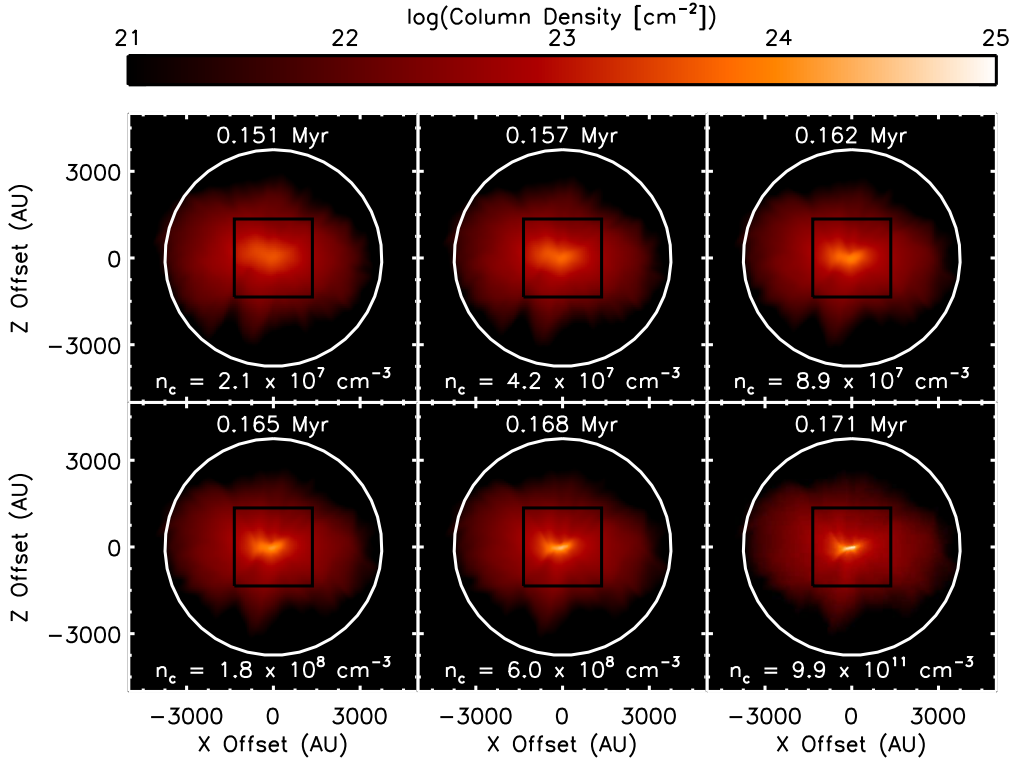}
\caption{\label{fig_sim_n_lowm}Total gas column density snapshots for the last 
0.02 Myr of evolution for simulation C04.  Each panel is labeled with the 
time that has elapsed since the simulation began, and the central density 
of the core at that time.  The last panel at 0.171 Myr is when a sink particle 
is inserted and the first hydrostatic core stage begins.  The white circles 
show the ALMA 106 GHz primary beam assuming a distance of 150 pc, and the 
black squares show the central region displayed in Figure 
\ref{fig_sim_alma_lowm}.}
\end{figure*}

\begin{figure*}[hbtp]
\epsscale{0.94}
\plotone{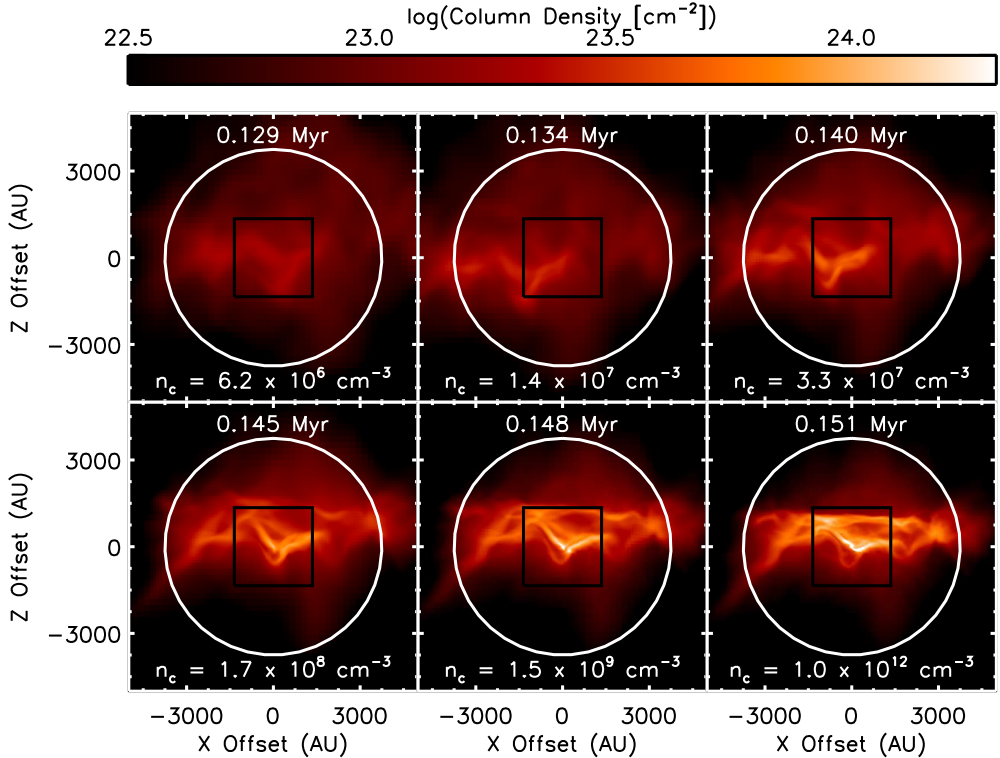}
\caption{\label{fig_sim_n_highm}Same as Figure \ref{fig_sim_n_lowm}, except 
for the last 0.022 Myr of evolution of the C4 simulation.  The black squares 
show the central region displayed in Figure \ref{fig_sim_alma_highm}.}
\end{figure*}

\begin{figure*}[hbtp]
\epsscale{0.92}
\plotone{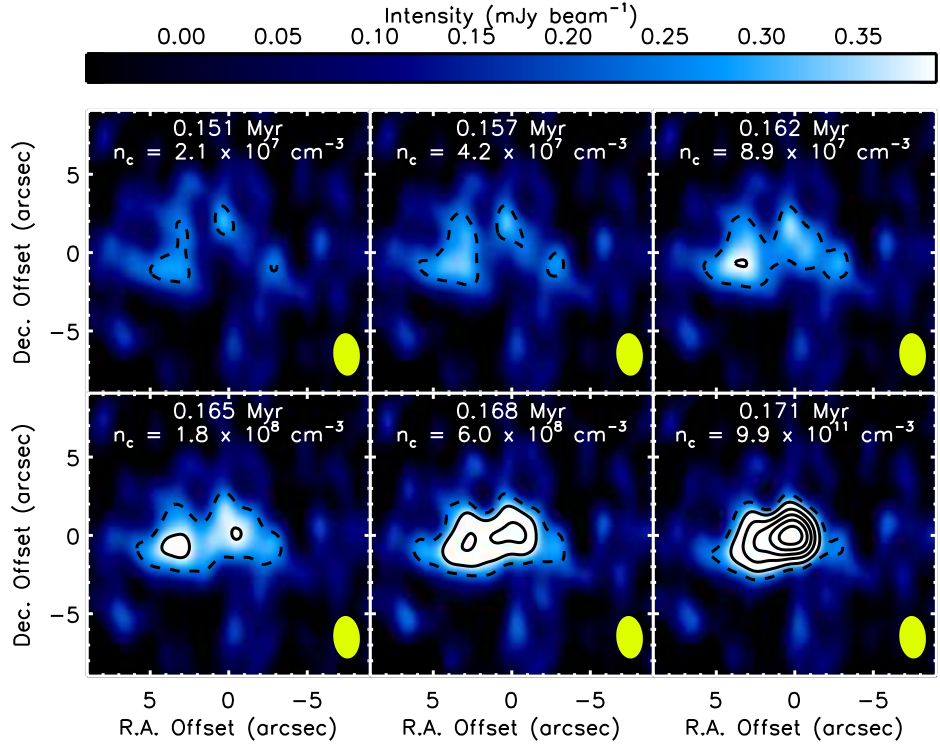}
\caption{\label{fig_sim_alma_lowm}Synthetic 106 GHz continuum observations of 
the C04 simulation at the same times shown above in Figure 
\ref{fig_sim_n_lowm}, matching the sensitivity and uv coverage of our ALMA 
Cycle 1 observations.  Each panel is labeled with the time that has elapsed 
since the simulation began, and the central density of the core at that time.  
The beam is shown by the yellow ellipse at the lower 
right of each panel.  The solid contours start at 5$\sigma$ and increase by 
2$\sigma$, where 1$\sigma$ $\sim$ 0.1 mJy beam$^{-1}$, matching the 
sensitivity of our observations (see Table \ref{tab_observations}).  The 
dashed contour plots the 3$\sigma$ level and is plotted as dashed 
to emphasize that it does not represent a robust detection.}
\end{figure*}

\begin{figure*}[hbtp]
\epsscale{0.92}
\plotone{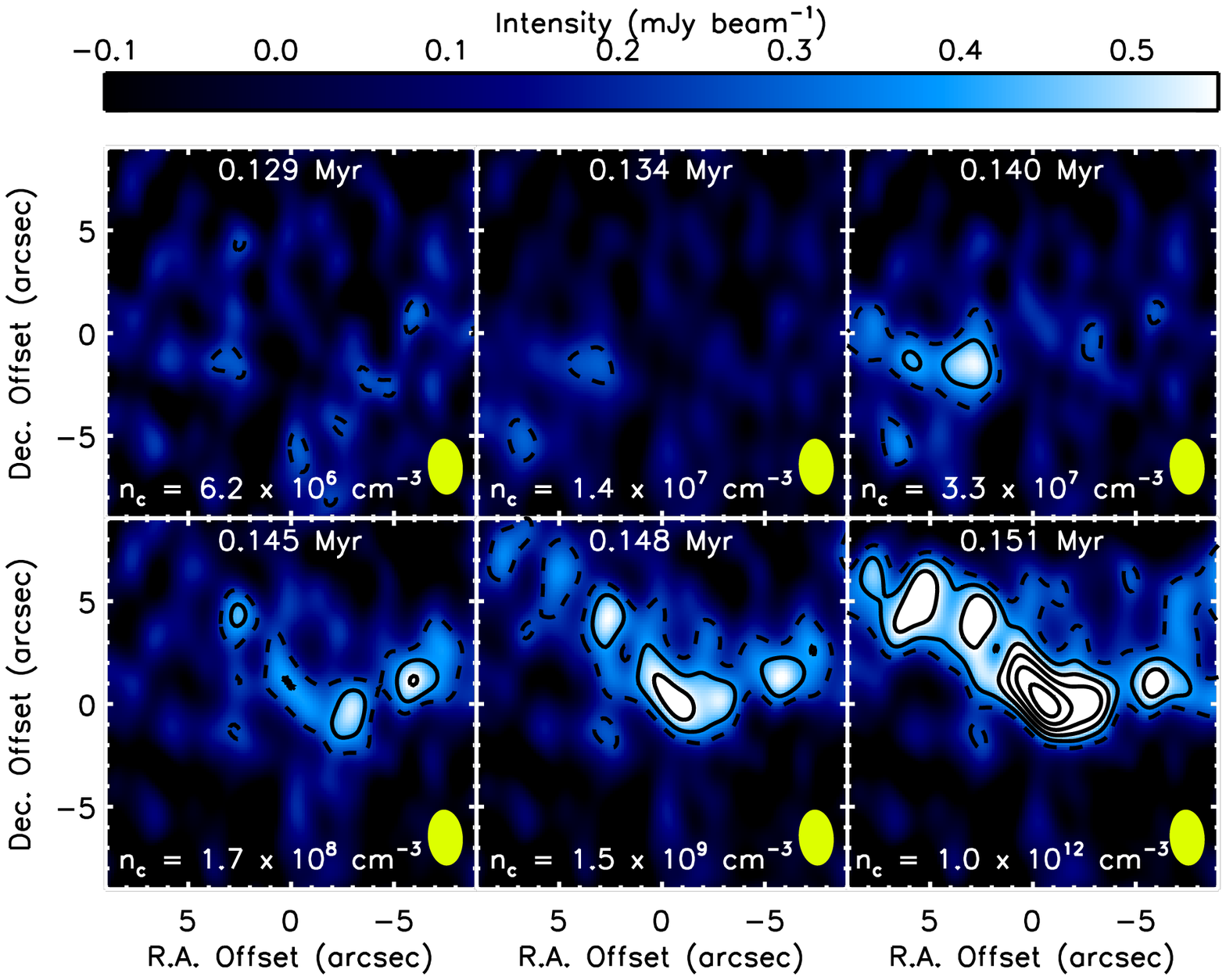}
\caption{\label{fig_sim_alma_highm}Same as Figure \ref{fig_sim_alma_lowm}, 
except for the last 0.022 Myr of evolution of the C4 simulation.}
\end{figure*}

The initial basegrid resolution is $64^3$, but the dense core is refined by 
two additional levels via a density threshold refinement criterion. As the 
gas collapses, additional AMR levels are inserted when the density exceeds the 
Jeans condition for a Jeans number of $N_J=0.0625$ \citep{truelove97}. Each 
subsequent level increases the cell resolution by a factor of 2. When the 
Jeans condition is violated on the eighth level, a sink particle is added, 
representing the formation of a protostar \citep{krumholz04}. We evolve the 
calculation until shortly after the formation of a sink particle. At that time, 
the gas densities are comparable to those expected for the formation of a 
first hydrostatic core 
\citep[$\rho= 10^{-12}$ g~cm$^{-3}$; see, e.g.,][]{tomida13}.

Figures \ref{fig_sim_n_lowm} and \ref{fig_sim_n_highm} show snapshots of the 
total gas column densities, $N$, from the two simulations, viewed 
perpendicular to the magnetic field (which is oriented in the $z$-direction).  
These snapshots focus on the last $\sim$0.02 Myr of evolution before 
the insertion of sink particles and onset of the first hydrostatic core 
stage.  Since both simulations have the same number of AMR levels, the C04 
core reaches higher densities before sink particle formation.

\subsubsection{Description of the Synthetic Observations}\label{sec_synthetic}

To generate synthetic ALMA observations matching our Cycle 1 observations of 
the dense cores in Chamaeleon~I, we begin with the total gas column density 
snapshots shown above.  These snapshots are output on a fixed grid with 
a pixel size of 20 AU (0.13\arcsec\ at the assumed distance of 150 pc), and 
converted to total gas surface density $\Sigma = \mu m_H N$, 
where $\mu$ and $m_H$ are the same as in Equation 
\ref{eq_numberdens}.  While the Chamaeleon~I dense cores are embedded within 
the larger cloud complex, the simulations are of isolated starless cores.  
This difference could affect our results since the cloud material will decrease 
the surface density contrast between the center and edge of the core.  To 
account for this difference, we add a constant value of $\Sigma$ representing 
the background cloud material to every pixel.  We consider the most extreme 
case by adopting a value of $\Sigma$ representing the maximum visual extinction 
toward the Chamaeleon~I cloud of $A_{V} = 35.6$, 
as measured in extinction maps with 2\arcmin\ resolution.  
These maps were produced
by the {\it Spitzer} c2d \citep{evans2009:c2d} and Gould Belt
 \citep{dunham2015:gb} survey teams using 2MASS \citep{skrutskie2006:2mass} 
and \spitzer\ data \citep[see][for details]{heiderman2010:threshold}.  We 
convert this visual extinction to hydrogen column density using the relation 
$N_H/A_V = 1.37 \times 10^{21}$~cm$^{-2}$ 
\citep[see][for details]{evans2009:c2d}, giving 
$N_H = 4.88 \times 10^{22}$~cm$^{-2}$.  The total gas surface density is then 
calculated as $\Sigma = 0.5 \mu_{H_2} m_H N_H$, where $\mu_{H_2}$ is the mean 
molecular weight per hydrogen molecule 
\citep[$\mu_{H_2} = 2.8$;][]{kauffmann2008:mambo} and the factor of 0.5 
accounts for the number of hydrogen atoms per hydrogen molecule, giving 
$\Sigma = 0.114$~g~cm$^{-2}$.  Later tests showed that proceeding without 
adding this offset representing the large-scale cloud material gives 
identical results, as all of the large-scale emission is filtered out by the 
interferometer.

With the mass in each pixel calculated as 
$M = \Sigma A_{\rm pixel}$, where $A_{\rm pixel}$ is the area of each pixel 
($A_{\rm pixel} = 400$ ${\rm AU}^2 = 9.0 \times 10^{28}$~${\rm cm}^{2}$), we 
then generate 106 GHz continuum maps by inverting Equation 
\ref{eq_dustmass} for flux density.   For the temperature in each pixel we use 
empirically derived relationships between temperature and surface density 
derived from models of Bonnor-Ebert spheres 
\citep{bonnor1956:bespheres,ebert1955:bespheres}, as described in Appendix 
A.  We use the same OH5 dust opacity as that used in Equation 
\ref{eq_dustmass} (0.23 cm$^2$~g$^{-1}$; see the text after Equation 
\ref{eq_dustmass} for a discussion of uncertainties in this assumed opacity).
%As noted above, the 3 mm dust 
%opacity may be up to an order of magnitude larger if dust grains have grown 
%to millimeter sizes or larger.  While growth to such large grains is 
%unexpected at the scales of filaments and cores, some recent work has 
%found evidence for such growth \citep{schnee2014:beta}. Increasing the 
%opacities by one order of magnitude would increase the 106 GHz continuum 
%intensities by the same amount, increasing the detectability of the 
%simulations and models considered in this paper.  If such grain growth 
%at the scale of dense cores is confirmed by future studies, the results 
%of this paper will require revisiting.

We then generate synthetic ALMA observations of these simulated 106 GHz 
continuum intensity maps using the CASA tasks {\sc simobserve} and 
{\sc simanalyze}.  We set the coordinates of the image centers to 
R.A.~=~11:00:00, decl.~=~$-$77:20:00, representing the approximate center 
of the Chamaeleon~I cloud.  We use 1 s integration times and include 
thermal noise from the atmosphere.  Matching our actual observations, for 
each core we observe a single pointing in the ALMA Cycle 1 configuration 
contained in the CASA configuration file {\sc alma.cycle1.3.cfg} for a total 
of 72 s (on-source), at a central frequency of 106 GHz with a total bandwidth 
of 6 GHz.  We then invert into the image plane and clean to a threshold of 
0.3~mJy (approximately 3$\sigma$) using non-interactive cleaning with no 
specified clean mask.

Figures \ref{fig_sim_alma_lowm} and \ref{fig_sim_alma_highm} show the resulting 
synthetic ALMA observations for the C04 and C4 simulations, respectively, 
at the same timesteps as shown above in the column density images.  As with 
the column density snapshots shown above, these images are for cores viewed 
perpendicular to the magnetic field (which is oriented in the $z$-direction).  
For each simulation, the third panel (top right panel) shows the first timestep 
where the core is detected to at least 5$\sigma$, and the last panel (bottom 
right panel) shows the onset of the first hydrostatic core stage.  
The C04 simulation is detected over the last 8200 yr of the starless 
stage, when the central number density exceeds $8.9 \times 10^{7}$~cm$^{-3}$.  
The C4 simulation is detected over the last 11000 yr of the starless stage, 
when the central number density exceeds $3.3 \times 10^{7}$~cm$^{-3}$.  
We emphasize here that these detection thresholds are for cores viewed 
perpendicular to the magnetic field axis; the effects of viewing geometry on 
our results will be discussed below in \S \ref{sec_detecting}.

\begin{figure}
%[hbtp]
\epsscale{1.2}
\plotone{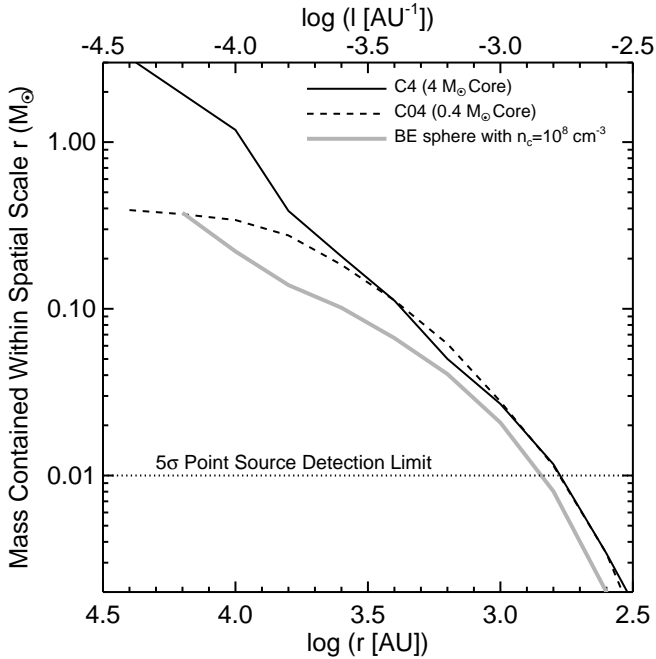}
\caption{\label{fig_fft}Mass within various spatial scales versus those 
spatial scales for the 
C4 (solid black line) and C04 (dashed black line) simulations, 
calculated at the first timestep each simulation is detected in our synthetic 
ALMA observations.  Also shown are the results for a Bonnor-Ebert sphere with a 
central density of $10^8$~cm$^{-3}$ (solid gray line; see Appendix A for 
details).  The masses are calculated by Fourier transforming images of the 
mass in each pixel obtained from the surface density maps as described in 
the text.  The dotted horizontal line shows the approximate 5$\sigma$ 
detection limit for point sources based on the sensitivities quoted in Table 
\ref{tab_observations}.}
\end{figure}

Initially, it seems surprising that the central density thresholds for 
detection are similar despite a factor of ten difference in initial core 
mass.  To further examine this situation, 
Figure \ref{fig_fft} plots, at the first detected timestep for 
each simulation, the mass contained within given spatial scales.  We calculate 
this by Fourier transforming images of the mass in each pixel calculated 
directly from the surface density maps, as described above, and then plotting 
the amplitude as a function of both spatial frequency $l$ (top axis) and 
spatial scale $r$ ($r=1/l$; bottom axis).  This method mimics what an 
interferometer like ALMA actually sees, and shows that the amount of mass on 
compact ($\sim 1000$ AU) scales is similar between the two simulations 
even though the total mass on 
larger scales is different.  The simulations considered here form compact 
structures with similar total masses when the cores reach similar central 
densities, regardless of the total mass reservoirs available in the cores.

\subsection{Detecting Starless Cores}\label{sec_detecting}

Based on the results of the previous section, starless cores evolving under 
the \citet{offner2010:turbfrag} turbulent fragmentation scenario should 
be detectable by our ALMA observations when they are viewed perpendicular to 
their magnetic field axes and their central number densities exceed between 
approximately 
$10^7$~cm$^{-3}$ and $10^8$~cm$^{-3}$, with the exact value depending on 
the mass of the core.  As the mean mass of the starless cores in Chamaeleon~I 
is 0.3 \msun\ \citep{belloche2011:chami}, we consider the C04 simulation to be 
more representative of Chamaeleon~I cores.  Thus, starless cores viewed 
perpendicular to their magnetic field axes should be detectable when their 
central number densities exceed $8.9 \times 10^{7}$~cm$^{-3}$.

While we do not have any way to measure the central number densities of the 
starless cores in Chamaeleon~I, 
\citet{belloche2011:chami} derived lower limits by converting the 
single-dish LABOCA 870 \um\ peak intensities to peak masses according to 
Equation \ref{eq_dustmass}, and then calculating peak number densities using 
the effective radius of the beam.  Their resulting values are 
$2.8-14.4~\times~10^5$~cm$^{-3}$, with a mean of $4.3 \times 10^5$~cm$^{-3}$.  
As the peak intensities are beam-diluted by the LABOCA beam (FWHM of 
21.2\arcsec, or 3180 AU at the distance to Chamaeleon I), the resulting 
peak number densities are lower limits to the 
true central number densities.  Indeed, to demonstrate this, we 
generated 870 \um\ images of the simulations following the same procedure as 
described above for the 106 GHz (2.8 mm) images, convolved them with the 
21.2\arcsec\ FWHM LABOCA beam, and calculated peak number densities as 
described above, finding that the calculated peak number density remains 
constant at a few~$\times 10^5$~cm$^{-3}$ as the central density increases from 
$10^7-10^{12}$~cm$^{-3}$.  Thus we take the minimum calculated peak number 
density of $2.8~\times~10^5$~cm$^{-3}$ as a strong lower limit to the 
central number densities of the 56 starless cores in Chamaeleon~I.

If we assume that Chamaeleon~I has experienced a continuous rate of star 
formation over a timescale at least as long as the core lifetimes $\tau$, which 
are approximately $10^6$~yr (see below), then the expected number of 
detections can be calculated as:

\begin{equation}\label{eq_detect1}
{\rm Detections} = {\rm N_{total}} \times \frac{\tau_{\rm Detectable}}{\tau_{\rm Total}} \qquad ,
\end{equation}
where ${\rm N_{total}}$ is the total number of cores observed, 
$\tau_{\rm Detectable}$ is the lifetime over which cores are detectable, and 
$\tau_{\rm Total}$ is the total core lifetime.  
If we further assume that the cores evolve on free-fall timescales such that 
their lifetimes as a function of central number densities scale as 
$\tau \propto n^{-0.5}$, then Equation \ref{eq_detect1} can be written as:

\begin{equation}\label{eq_detect2}
{\rm Detections} > {\rm N_{total}} \times \left( \frac{n_{\rm Detectable}}{n_{\rm Limit}} \right)^{-0.5} \qquad ,
\end{equation}
where $n_{\rm Detectable}$ is the central density threshold for detection and 
$n_{\rm Limit}$ is the observed lower limit for the central number densities of 
the cores.  Note that the equation becomes an inequality since $n_{\rm Limit}$ 
is a lower limit.  With N$_{\rm total} = 56$, 
n$_{\rm Detectable} = 8.9 \times 10^7$~cm$^{-3}$, and 
n$_{\rm Limit} > 2.8 \times 10^5$~cm$^{-3}$, Equation \ref{eq_detect2} 
results in a total of at least three expected detections.

\begin{figure*}[t]
\epsscale{1.0}
\plotone{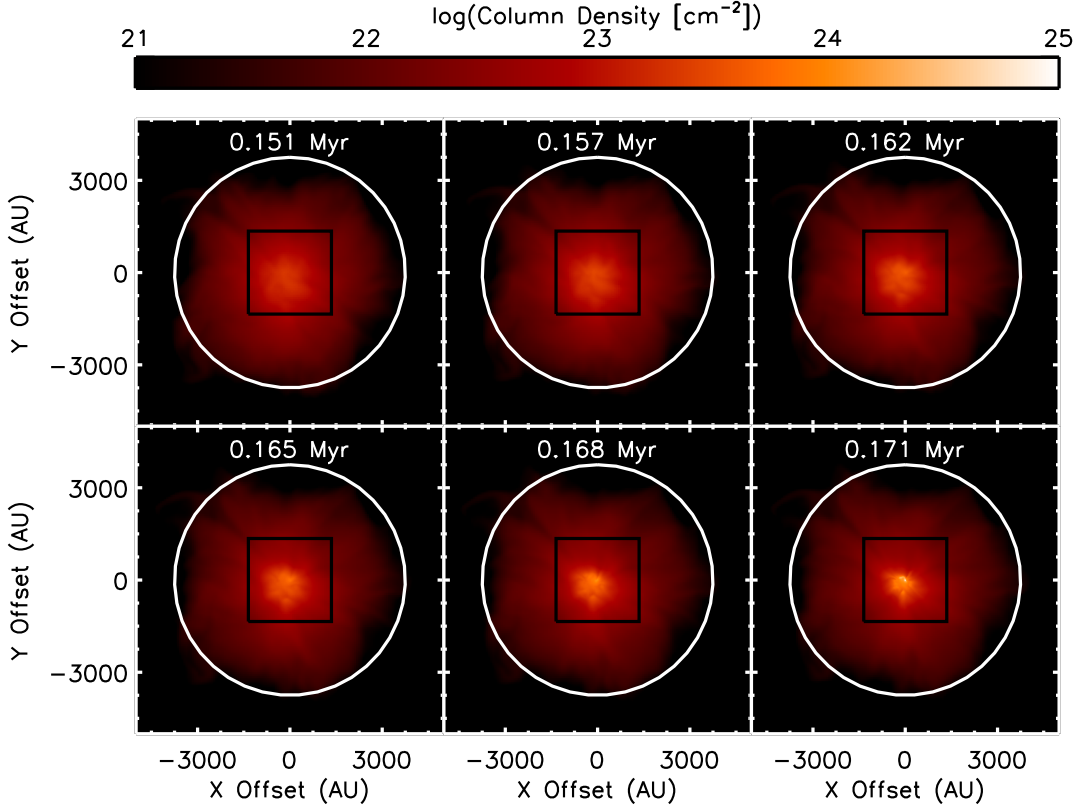}
\caption{\label{fig_sim_n_lowm_zproj}Same as Figure \ref{fig_sim_n_lowm}, 
except viewed down the magnetic field axis.}
\end{figure*}  

As noted above, these results are derived using synthetic observations of 
simulated cores viewed perpendicular to their magnetic field axes.  Since the 
cores tend to collapse along the magnetic field lines, they develop elongated, 
filamentary structure as they collapse.  With this viewing geometry the 
elongated structure is seen edge-on, increasing the maximum column density 
observed at a given timestep.  Indeed, Figure \ref{fig_sim_n_lowm_zproj} shows 
column density snapshots for the C04 simulation at the same timesteps as shown 
in Figure \ref{fig_sim_n_lowm}, except viewed down the magnetic field axis 
(i.e., rotated 90\degree\ compared to Figure \ref{fig_sim_n_lowm}).  
With this viewing geometry the cores exhibit less filamentary structure and 
lower column densities.  After averaging over all possible orientations 
with respect to the line of sight, the total number of expected detections 
is reduced from at least three to at least two.

\subsection{Implications of the 56 Non-Detections}

Assuming Poisson statistics, the probability of detecting zero cores when 
two detections are 
expected is 13.5\%.  While there is thus a non-negligible 
chance that the non-detections are simply due to the sample size being too 
small, it is more likely that at least one of the assumptions 
adopted above are wrong. Here we evaluate each 
of these three assumptions, namely:  (1) starless cores evolve on timescales 
proportional to the free-fall time, (2) star formation is continuous in 
Chamaeleon~I, and (3) the simulations are applicable to the starless cores 
in Chamaeleon~I.

\subsubsection{Do Starless Cores Evolve on Timescales Proportional to the Free-Fall Time?}

\begin{figure}
%[hbtp]
\epsscale{1.2}
\plotone{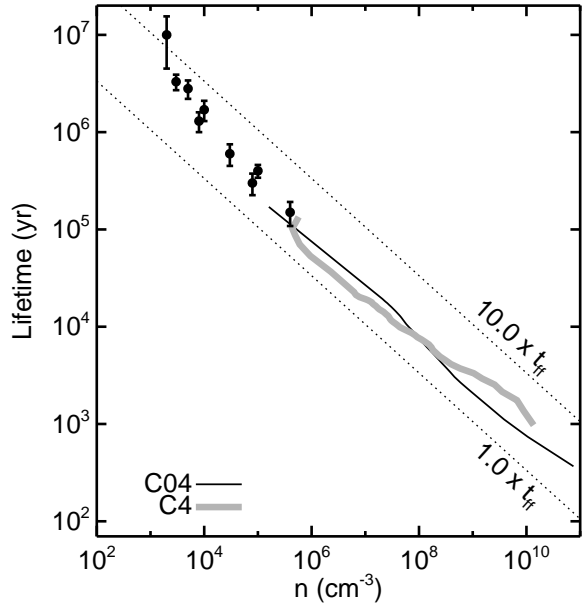}
\caption{\label{fig_lifetimes}Lifetimes versus central number densities for 
the C04 (thin black line) and C4 (thick gray line) simulations as they 
evolve.  The lifetime at a given density is calculated as the amount of time 
it takes the simulation to evolve from the timestep with that central density 
to the onset of the first hydrostatic core phase.  The dashed lines show 
the free-fall lifetime ($t_{\rm ff} = \sqrt{(3 \pi)/(32 G \rho)}$) and 
$10 t_{\rm ff}$.  The data points are taken from the literature as compiled by 
\citet{jessop2000:starless} and \citet{wardthompson2007:ppv}.}
\end{figure}

In order to evaluate whether or not 
starless cores evolve on timescales proportional 
to the free-fall time, meaning $\tau \propto n^{-0.5}$, Figure 
\ref{fig_lifetimes} plots the lifetime versus central number density 
\citep[a so-called `JWT' plot;][]{jessop2000:starless} for the 
C04 and C4 simulations.  The lifetime at a given density is calculated as the 
amount of time it takes the simulation to evolve from the timestep with that 
central density to the onset of the first hydrostatic core phase.  The C04 
simulation, which is most appropriate for comparison to the cores in 
Chamaeleon~I given their similar masses, lies between the $t_{\rm ff}$ and 
$10 t_{\rm ff}$ lines but follows an identical slope, with a power-law 
fit giving $\tau~\propto~n^{-0.50 \pm 0.01}$.  A power-law fit to the C4 
simulation gives $\tau~\propto~n^{-0.41 \pm 0.01}$.

Also plotted in Figure \ref{fig_lifetimes} are observational determinations 
of $\tau$ vs.~central number density for various samples from the literature, 
as compiled by \citet{jessop2000:starless} and \citet{wardthompson2007:ppv}.  
Visual inspection of the data suggests that core lifetimes normalized to 
free-fall times decrease as cores evolve to higher central densities, and 
indeed a power-law fit to all available data gives 
$\tau~\propto~n^{-0.69 \pm 0.06}$.  However, restricting the fit only to 
starless core populations with densities above $10^4$~cm$^{-3}$, densities 
more relevant for comparison to the high densities at which cores become 
detectable by our ALMA observations, the data follow the power-law 
$\tau~\propto~n^{-0.53 \pm 0.11}$, consistent to within 1$\sigma$ with our 
assumption of evolution proportional to free-fall times.  Thus we conclude 
that there is both observational and theoretical support for our 
assumption that starless cores evolve on timescales proportional 
to the free-fall timescale.

We note that, very recently, \citet{konyves2015:herschel} 
found evidence that starless cores in Aquila evolve on faster timescales.  
While they don't report the results of a power-law fit to 
$\tau$~vs.~$n$, we estimate from their Figure 9 that the index of such a 
power-law is $\sim -0.9$ for central densities between $10^5$~cm$^{-3}$ and 
$10^6$~cm$^{-3}$.  Inserting such a power-law index in Equation 
\ref{eq_detect2} would give much less than one expected detection.  
If future studies confirm the \citet{konyves2015:herschel} results in 
additional clouds, then our assumption that $\tau \propto n^{-0.5}$ will 
require revisiting.

\subsubsection{Is Star Formation Continuous in Chamaeleon~I?}

\begin{figure}
%[hbtp]
\epsscale{1.2}
\plotone{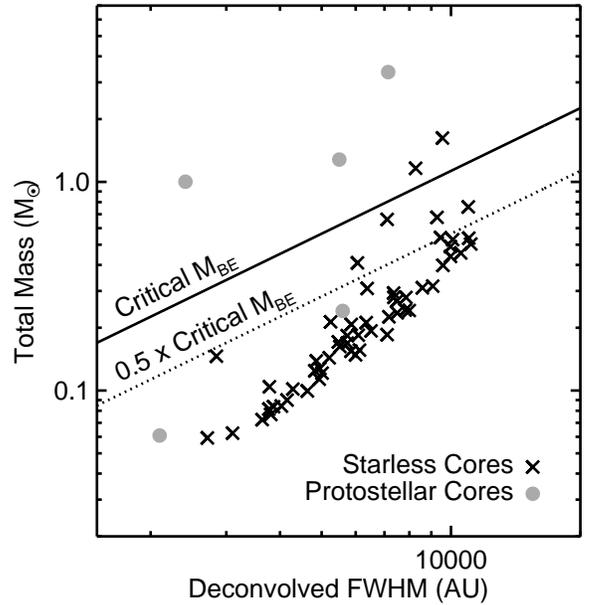}
\caption{\label{fig_mass_size}Mass vs.~size (deconvolved FWHM) for the 
starless (black x symbols) and protostellar (gray circles) cores in 
Chamaeleon~I, taken from \citet{belloche2011:chami}.  The solid line 
shows the critical Bonnor-Ebert mass ($M_{\rm BE}$), the maximum mass at a 
given radius for which a Bonnor-Ebert sphere is stable.  The dotted line 
shows $0.5 \times M_{\rm BE}$.}
\end{figure}

The assumption of continuous star formation means that starless cores form and 
collapse into stars at a constant rate.  To test this assumption, 
we first examine the ratio of Class 0+I to Class II Young Stellar Objects 
in Chamaeleon~I compared to other star-forming clouds.  According to the 
results compiled by \citet{dunham2015:gb}, this ratio is 0.06 for Chamaeleon~I 
compared to 0.32 averaged over the 18 clouds observed by the {\it Spitzer} 
c2d \citep{evans2009:c2d} and Gould Belt \citep{dunham2015:gb} Legacy 
Surveys.  Thus, Chamaeleon~I is deficient in the fraction of the youngest 
protostars compared to other, nearby molecular clouds, possibly indicating a 
decelerating rate of star formation, a point also noted by 
\citet{belloche2011:chami}.

To further test the assumption of continuous star formation, Figure 
\ref{fig_mass_size} plots the mass versus size for the starless and 
protostellar cores in Chamaeleon~I, as tabulated by 
\citet{belloche2011:chami}.  Overplotted are lines showing the maximum mass 
at a given radius for which a Bonnor-Ebert sphere 
\citep{bonnor1956:bespheres,ebert1955:bespheres} is stable, and 
0.5 times this maximum mass (to account for the typical factor of $\sim$2 
uncertainty in mass calculations).  \citet{belloche2011:chami} present a 
very similar figure (their Figure 9b) and argue that most of the starless 
cores in Chamaeleon~I are in fact stable and not currently collapsing, and 
thus not evolving to form stars at a constant rate.

\citet{belloche2011:chami} combine the above two arguments - namely a 
low protostar fraction and low fraction of unstable starless cores - with 
a measured high global star formation efficiency to argue that Chamaeleon~I 
has a declining star formation rate and is nearly finished forming stars.  
Such a scenario may be explained if Chamaeleon~I is globally unbound 
due to turbulence, as such a cloud would produce a population of starless 
cores that would not collapse further \citep{clark2004:unbound}.  
In this scenario, most of the remaining starless cores will never form stars.  
While \citet{naranjoromero2015:starless} present a scenario in which 
collapsing starless cores can appear stable when compared to their critical 
Bonnor-Ebert masses, other authors find that applying the critical Bonnor-Ebert 
stability criterion overpredicts the number of unstable cores 
\citep[e.g.,][]{pattle2015:scuba2}.  Furthermore, \citet{tsitali2015:chami} 
found that only five of the starless cores in Chamaeleon~I (9\%) are bound, 
and only 13--28\% show kinematic signatures of infall.  
We note that many of these cores, which are typically located in 
large-scale, filamentary cloud structures, could accrete more mass and thus 
become unstable in the future.  
While \citet{belloche2011:chami} argue 
against this based on the low mass-to-length ratios of the filaments, 
the filaments themselves could still be gaining mass from the surrounding 
cloud.  Indeed, further analysis by \citet{belloche2011:chamiii} suggested 
that up to 50\% of the starless cores in Chamaeleon~I could eventually 
go on to form stars.

\begin{figure*}[hbtp]
\epsscale{0.92}
\plotone{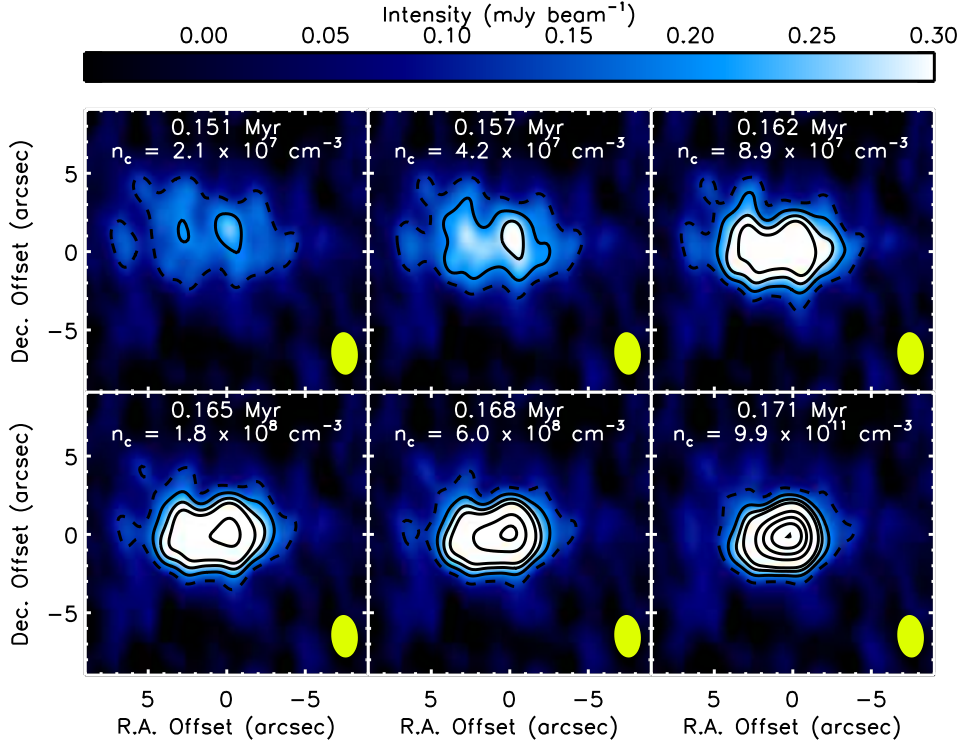}
\caption{\label{fig_sim_alma_lowm_4xlonger}Same as Figure 
\ref{fig_sim_alma_lowm}, except with four times longer total 
integration time.  The dashed contour 
plots the 3$\sigma$ level, where 1$\sigma$ $\sim$ 0.05 mJy beam$^{-1}$, and is 
plotted as dashed to emphasize that it does not represent a robust detection.  
The first three solid contours plot 5$\sigma$, 7$\sigma$, 
and 9$\sigma$ emission.  The remaining contours start at 15$\sigma$ and 
increase in increments of 5$\sigma$.}
\end{figure*}

Given the above evidence, the assumption of continuous star formation in 
Chamaeleon~I is likely wrong.  Instead, 
the star formation rate in this cloud is declining and most of the 
starless cores currently observed are not collapsing.  
Whether or not star formation is 
truly ending or simply pausing remains an open question.  Either way, 
since the simulations are collapsing at all times, N$_{\rm total}$ in Equations 
\ref{eq_detect1} and \ref{eq_detect2} should be the total number of 
\textit{collapsing} starless cores.  Reducing this number by 
more than a factor of 
two would decrease the lower limit to the number of expected detections to 
below one.  Thus our non-detections may be fully explained by a false 
assumption of continuous star formation.

\subsubsection{Are the Simulations Applicable?}

Even if all of the starless cores in Chamaeleon~I are collapsing, which 
evidence suggests is not the case (as discussed above), a third 
possibility is that the simulations themselves are not applicable to 
these starless cores in terms of their initial conditions or physical 
treatment.  
As noted in Table \ref{simprop}, the simulations start with 
initial velocity dispersions between 0.16--0.26 \kms.  These values are 
dominated by turbulent motions that are damped as the cores collapse, 
leading to lower values at the times the cores are actually detectable.  
For example, the velocity dispersion in the C04 simulation decreases from 
0.16 \kms\ at early times to $\sim$0.1~\kms\ just prior to the formation 
of the first core.  These velocity dispersions are in good agreement with 
those recently measured for the Chamaeleon~I starless cores by 
\citet{tsitali2015:chami}.  
However, it is still possible that the simulations are not applicable 
to the Chamaeleon~I starless cores.

As an alternative to the simulations considered here, 
Appendix A shows that starless cores collapsing as Bonnor-Ebert spheres 
\citep{bonnor1956:bespheres,ebert1955:bespheres} remain undetected to 
central densities at least as high as $10^{10}$~cm$^{-3}$.  Inserting 
$n_{\rm Detectable}~>~10^{10}$~cm$^{-3}$ in Equation \ref{eq_detect2} results in 
a lower limit to the number of expected detections of much less than one.  
While this could also explain our non-detections, we are unable to distinguish 
between the simulations and Bonnor-Ebert spheres due to the fact that most 
of the cores are likely not collapsing, as discussed above.

While our survey of the starless cores in Chamaeleon I is an important step 
toward understanding when substructure and fragmentation develop in collapsing 
molecular cloud cores, further work is needed on both the observational and 
theoretical fronts.  On the theoretical side, a full parameter space 
exploration is required to test how the development of substructure and 
fragmentation depends on initial core mass, turbulence, magnetic field 
strength, and viewing geometry.  

On the observational side, the first step required is improved statistics.  
If all of the cores are indeed collapsing, 
increasing the total number of cores observed by a factor of three would 
increase the number of expected detections from 
two to six, reducing the probability of zero detections from 
13.5\% to 0.2\% (assuming Poisson 
statistics).  Furthermore, as shown in Figure \ref{fig_sim_alma_lowm_4xlonger}, 
integrating to two times deeper mass sensitivity would reduce the central 
density detection threshold by approximately a factor of four, doubling the 
number of expected detections.  We thus recommend that future surveys 
of starless cores be conducted with ALMA, and that these surveys integrate to 
twice the mass sensitivity obtained here in order to maximize their 
statistical power. 
Finally, beyond simply improving the statistics, future surveys 
should also focus on surveying starless cores that are likely to be unstable 
and thus collapsing, and on characterizing variations in the 
levels of substructure and fragmentation with different core properties, 
including their mass and level of turbulence.  Only with such full parameter 
space study can different models of collapsing cores be distinguished.

\section{Summary}\label{sec_summary}

In this paper, we have presented the results from a Band 3 (3 mm), ALMA Cycle~1 
survey of 73 sources in Chamaeleon~I previously identified in a 
single-dish 870 \um\ continuum survey.  We summarize our main results as 
follows:

\begin{enumerate}
\item We detect a total of 24 106 GHz continuum sources in 19 different 
target fields, leaving 54 target fields with no detections.  We calculate 
and tabulate the effective radius, total gas mass, and mean total gas number 
density for each detected source.
\item We tabulate associations between our continuum detections and 
sources detected by LABOCA, YSOs detected by \herschel, and protostars 
detected by \spitzer.  We compile SEDs for each detection and use the 
results to assign a final evolutionary status to each of the 73 targets, 
identifying 56 targets as starless cores.
\item All previously known Class 0 and Class I protostars in Chamaeleon~I are 
detected in our continuum observations.  All other detections are associated 
with more evolved Flat-spectrum or Class II YSOs.
\item Very low luminosity protostars or first hydrostatic cores below 
the sensitivities of \spitzer\ and \herschel\ would have been detected to at 
least $5\sigma$ significance under even the most pessimistic assumptions, based 
on comparing to both simulations and to 
similar objects previously detected in the Perseus molecular 
cloud.  Our results thus indicate that the \spitzer+\herschel\ census of 
protostars in Chamaeleon~I is complete, with the rate at which protostellar 
cores have been misclassified as starless cores calculated as $<$1/56, 
or $<$2\%.
\item Synthetic observations of magneto-hydrodynamical simulations of 
collapsing starless cores 
following the turbulent fragmentation scenario are detectable 
by these ALMA observations, but only over relatively short time periods when 
their central densities exceed $\sim$10$^8$~cm$^{-3}$, with the exact density 
dependent on the viewing geometry.  Bonnor-Ebert spheres, on the other hand, 
remain undetected to central densities at least as high as $10^{10}$~cm$^{-3}$.
\item Assuming that the turbulent fragmentation simulations accurately 
describe the collapse of starless cores in Chamaeleon~I, that star formation 
is continuous and all of the starless cores in Chamaeleon~I are collapsing, 
and that cores evolve on timescales proportional to their free-fall times, our 
sample of 56 starless cores, all with central densities above 
$2.8~\times~10^5$~cm$^{-3}$, should have yielded at least 
two detections.
\item Assuming Poisson statistics, there is a 13.5\% probability of 
zero detections if the above assumptions are correct.  Thus, while it is 
possible that our non-detections result from a sample size that is 
too small, it is more likely that at least one of the above assumptions are 
wrong.  The most likely culprit is the assumption of continuous star formation; 
instead, evidence suggests that the star formation rate in 
Chamaeleon~I is declining and most of the starless cores are not currently 
collapsing.  It is also possible that the cores are more accurately described 
by models that develop less substructure than the simulations considered here, 
such as Bonnor-Ebert spheres.
\end{enumerate}

With the advent of ALMA, surveys of the complete populations of starless 
cores in molecular clouds with sufficient sensitivity to detect substructure 
and fragments have now become accessible.  We have demonstrated the 
feasibility and usefulness of such surveys with this work targeting the 
dense core population of Chamaeleon~I.  Future work should target additional 
clouds in order to improve the statistics and better constrain the validity 
of the turbulent fragmentation scenario in the evolution of starless cores 
and origin of multiple systems.  Surveys of additional clouds are also needed 
to fully determine the fraction of protostellar cores that have been 
misclassified as starless given the vastly different fractions determined 
here for Chamaeleon~I ($<$2\%) and previously by \citet{schnee2012:starless} 
for Perseus (possibly as high as 20\%).  

\acknowledgments
We thank the referee for a set of helpful comments that have 
improved the quality of this publication.  
The National Radio Astronomy Observatory is a facility of the National
Science Foundation operated under cooperative agreement by Associated
Universities, Inc.  This paper makes use of the following ALMA data:
ADS/JAO.ALMA\# 2012.1.00031.S. ALMA is a partnership of ESO
(representing its member states), NSF (USA) and NINS (Japan), together
with NRC (Canada) and NSC and ASIAA (Taiwan), in cooperation with the
Republic of Chile. The Joint ALMA Observatory is operated by ESO,
AUI/NRAO and NAOJ.
This research has made use of data from the Herschel Gould Belt survey (HGBS) 
project (http://gouldbelt-herschel.cea.fr). The HGBS is a Herschel Key 
Program jointly carried out by SPIRE Specialist Astronomy Group 3 (SAG 3), 
scientists of several institutes in the PACS Consortium (CEA Saclay, INAF-IFSI 
Rome and INAF-Arcetri, KU Leuven, MPIA Heidelberg), and scientists of the 
Herschel Science Center (HSC).
This publication makes use of data products from the Two Micron All Sky Survey, 
which is a joint project of the University of Massachusetts and 
from the Infrared Processing and Analysis Center/California 
Institute of Technology, funded by the National Aeronautics and Space 
Administration and the National Science Foundation.  These data were provided 
by the NASA/IPAC Infrared Science Archive, which is operated by the Jet 
Propulsion Laboratory, California Institute of Technology, under contract with 
NASA.  
This research has also made use of NASA's Astrophysics Data System (ADS) 
Abstract Service, the IDL Astronomy Library hosted by the NASA Goddard Space 
Flight Center, and the SIMBAD database operated at CDS, Strasbourg, France.
M.~M.~D.~acknowledges support from the Submillimeter Array (SMA) through an 
SMA postdoctoral fellowship, and from NASA through grant NNX13AE54G.  
X.~C.~acknowledges support by the NSFC through grant 11473069.

\bibliographystyle{apj.bst}
\bibliography{dunham_citations,outflowbib}

\appendix

\section{A.~~Bonnor-Ebert Spheres}

In this appendix we assess the detectability of Bonnor-Ebert spheres 
\citep[pressure-truncated, self-gravitating spheres;][]{bonnor1956:bespheres,ebert1955:bespheres} in our ALMA observations.  
We constructed Bonnor-Ebert (hereafter BE) density profiles for 
central densities ranging from $10^5$~cm$^{-3}$ to $10^{10}$~cm$^{-3}$, 
truncated at 7000 AU to approximately match the mean FWHM of 6600 AU for the 
starless cores in Chamaeleon~I, based on the sizes measured by 
\citet{belloche2011:chami}.  We then solved for the temperature profiles in 
each BE sphere using the three-dimensional Monte Carlo radiative transfer 
package RADMC-3D\footnote{Available at: http://www.ita.uni-heidelberg.de/$\sim$dullemond/software/radmc-3d/} \citep{dullemond2000:radmc,dullemond2004:radmc} 
in its 1-D, spherically symmetric mode.  The cores are heated externally by 
the interstellar radiation field (ISRF).  We adopt 
the \citet{black1994:isrf} ISRF, modified in the ultraviolet to reproduce 
the \citet{draine1978:isrf} ISRF, and then extincted by a given $A_V$ of 
dust with properties given by \citet{draine1984:dust} to simulate extinction 
by the surrounding lower density environment.  Further discussion of this 
adopted ISRF is given by \citet{evans2001:starless}.  For the radiative 
transfer calculations, we adopt the dust opacities of \citet{ossenkopf1994:oh5} 
appropriate for thin ice mantles after $10^5$ yr of coagulation at a gas 
density of $10^6$ cm$^{-3}$ (OH5 dust), and include isotropic scattering off 
dust grains.

\begin{figure*}[h]
\epsscale{1.2}
\plotone{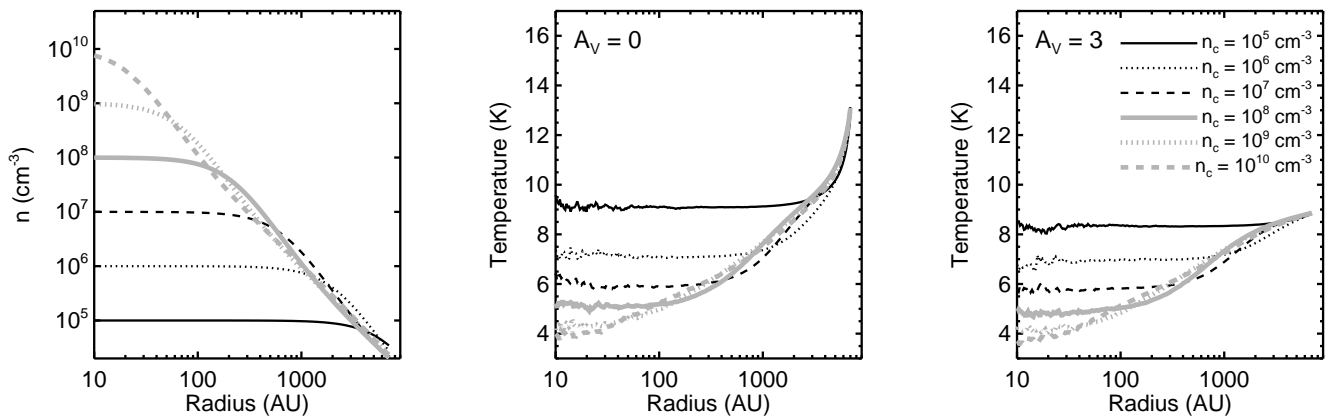}
\caption{\label{fig_be_profiles}Density and temperature profiles for the 
Bonnor-Ebert (BE) spheres considered here.  {\it Left:}  
Total gas number density profiles for BE spheres ranging in 
central densities from $10^5$~cm$^{-3}$ to 
$10^{10}$~cm$^{-3}$, truncated at 7000 AU.  {\it Middle:}  Temperature profiles 
calculated by RADMC-3D for the case of no attenuation of the interstellar 
radiation field (ISRF; see text for details).  {\it Right:}  Temperature 
profiles calculated by RADMC-3D for the case of attentuation of the ISRF 
by $A_V=3$ of dust with properties given by \citet{draine1984:dust}.  
The figure key given in the right panel applies to all three panels.}
\end{figure*}

\begin{figure*}[hbtp]
\epsscale{0.93}
\plotone{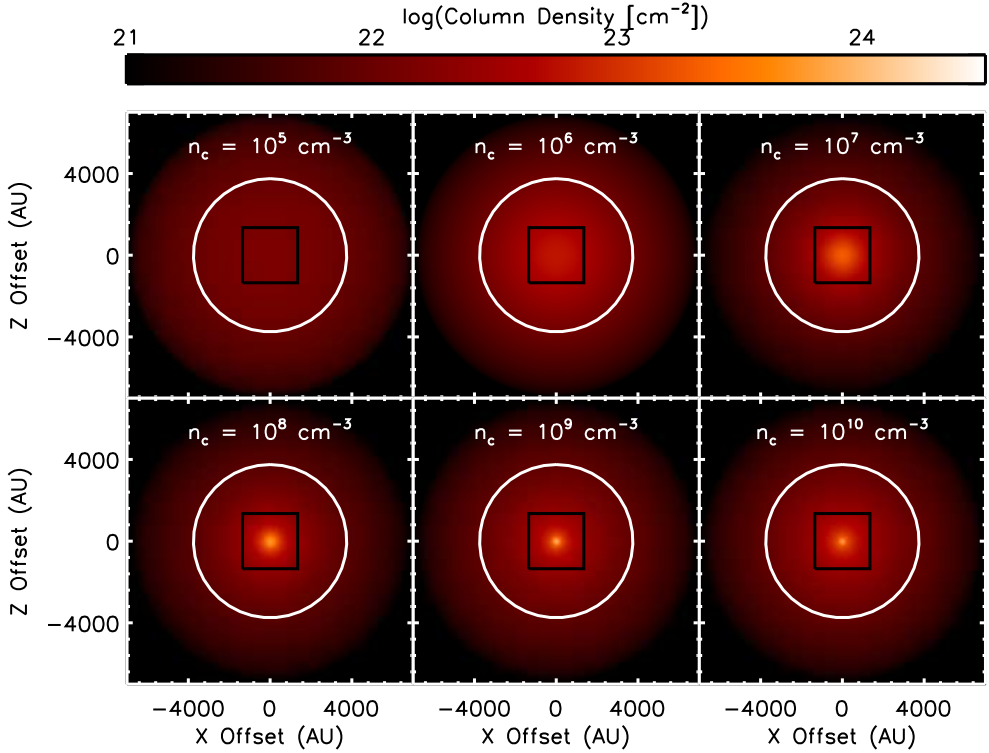}
\caption{\label{fig_sim_n_be}Total gas column density for Bonnor-Ebert 
spheres ranging in central density from $10^5$~cm$^{-3}$ to $10^{10}$~cm$^{-3}$ 
and truncated to radii of 7000 AU.  The white circles 
show the ALMA 106 GHz primary beam assuming a distance of 150 pc, and the 
black squares show the central region displayed in 
Figure~\ref{fig_sim_alma_be}.}
\end{figure*}

Figure \ref{fig_be_profiles} plots the number density and temperature profiles 
for these BE spheres, where the latter are calculated by RADMC-3D and 
agree with those calculated by \citet{evans2001:starless}.  Two 
different temperature profiles are shown, one set for no attenuation 
of the ISRF ($A_V=0$) and one set for attenuation of the ISRF by 
$A_V=3$ of dust.  As expected, the attenuated ISRF results in generally 
cooler temperature profiles, with less contrast between the dust temperatures 
at the centers and the outer edges.  For all of the following analysis, we 
consider only the case with $A_V=3$; identical results on detectability 
were obtained with the $A_V=0$ case.  Figure \ref{fig_sim_n_be} shows 
images of the total gas column density for the six different BE spheres 
considered here.

\begin{figure*}[hbtp]
\epsscale{0.92}
\plotone{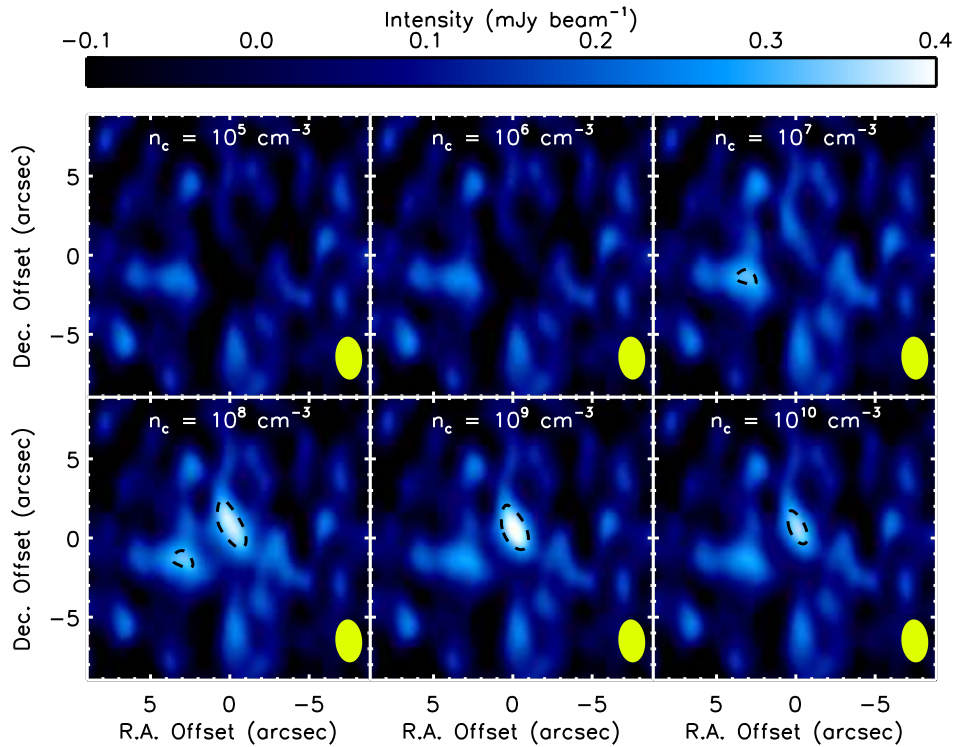}
\caption{\label{fig_sim_alma_be}Synthetic 106 GHz continuum observations of 
Bonnor-Ebert spheres ranging in central density from $10^5$~cm$^{-3}$ to 
$10^{10}$~cm$^{-3}$ and truncated to radii of 7000 AU, matching the sensitivity 
and uv coverage of our ALMA Cycle 1 observations.  The beam is shown by the 
yellow ellipse at the lower right of each panel.  The solid contours start at 
5$\sigma$ and increase by 2$\sigma$, where 1$\sigma$ $\sim$ 0.1 mJy 
beam$^{-1}$, matching the sensitivity of our observations (see Table 
\ref{tab_observations}).  The dotted contour plots the 3$\sigma$ level and is 
plotted as dashed to emphasize that it does not represent a robust detection.}
\end{figure*}

To generate synthetic ALMA observations matching our Cycle 1 observations 
of dense cores in Chamaeleon~I, we use RADMC-3D to generate model images at 
106 GHz.  We then generate synthetic ALMA observations of these model 
images following the same procedure as described above in \S 
\ref{sec_synthetic}: using the {\sc CASA} tasks {\sc simobserve} and 
{\sc simanalyze}, we set the image center to coordinates matching the 
approximate center of the Chamaeleon~I cloud and observe each model for a 
total of 72 s on-source in 1 second 
integrations with 6 GHz of bandwidth, using the 
ALMA Cycle 1 configuration contained in the {\sc CASA} configuration file 
{\sc alma.cycle1.3.cfg}.  We include thermal noise from the atmosphere and 
clean to a threshold of 0.3~mJy (approximately 3$\sigma$) using 
non-interactive cleaning with no specified clean mask.

Figure \ref{fig_sim_alma_be} shows the resulting synthetic ALMA observations 
for the six different BE spheres.  While there are 3$\sigma$ detections 
associated with the BE spheres with $n_c \geq 10^8$~cm$^{-3}$, 
there are no 5$\sigma$ detections, not even for the $n_c = 10^{10}$~cm$^{-3}$ 
model.  Thus, using the 5$\sigma$ detection threshold 
discussed above, Bonnor-Ebert spheres would remain undetected to central 
densities at least as high as $10^{10}$~cm$^{-3}$.  

Inserting $n_{\rm Detectable}~>~10^{10}$~cm$^{-3}$ into Equation \ref{eq_detect2} 
while leaving $N_{\rm total}~=~56$ and 
$n_{\rm Limit}~>~2.8~\times~10^5$~cm$^{-3}$ 
yields an expectation of less than one detection.  To understand these 
results, we note that the masses on various spatial scales shown in 
Figure \ref{fig_fft}, calculated via Fourier transforms of images of the 
mass in each pixel as described above, show that a BE sphere with a 
very similar central density to that of the C04 simulation when it is 
first detected ($10^8$~cm$^{-3}$ for the BE sphere versus 
$8.9~\times~10^7$~cm$^{-3}$ for the C04 simulation) has less mass on scales 
of several hundred AU to several thousand AU despite comparable total masses.  
Future surveys targeting larger numbers of starless cores in different 
environments are needed to fully distinguish between the levels of substructure 
and fragmentation predicted by Bonnor-Ebert spheres and those predicted 
by the \citet{offner2010:turbfrag} turbulent fragmentation simulations.

Inspection of Figure \ref{fig_be_profiles} shows that our model 
Bonnor-Ebert spheres reach very cold temperatures in their centers, as low 
as $4-5$ K for the models with $n_c \ge 10^{8}$~cm$^{-3}$.  While the BE sphere 
models presented by \citet{evans2001:starless} only reached minimum central 
temperatures of $6-7$ K, they only considered models with central densities 
up to $10^7$~cm$^{-3}$, thus our results are in fact consistent.  We 
acknowledge that our models only consider external heating from the 
interstellar radiation field while neglecting other sources of heating, 
including cosmic ray heating.  By setting a temperature floor of 7 K and then 
regenerating the synthetic ALMA images shown in Figure 
\ref{fig_sim_alma_be}, we find that the Bonnor-Ebert spheres produce 
5$\sigma$ detections for $n_c > 10^{8}$~cm$^{-3}$.

However, while the neglected additional sources of heating 
could increase the central temperatures, \citet{evans2001:starless} showed 
that, for dust temperatures as low as at least 5 K, external heating from the 
interstellar radiation field dominated over the sum of direct heating of dust 
by cosmic rays, heating of dust by ultraviolet photons created following 
cosmic ray ionization of H$_2$, and heating of dust by collisions with gas 
heated by cosmic rays.  Thus, we argue that our low temperatures, and 
resulting predictions that BE spheres remain undetected to central 
densities at least as high as $10^{10}$~cm$^{-3}$, are likely realistic.

\begin{figure*}[h]
\epsscale{0.9}
\plotone{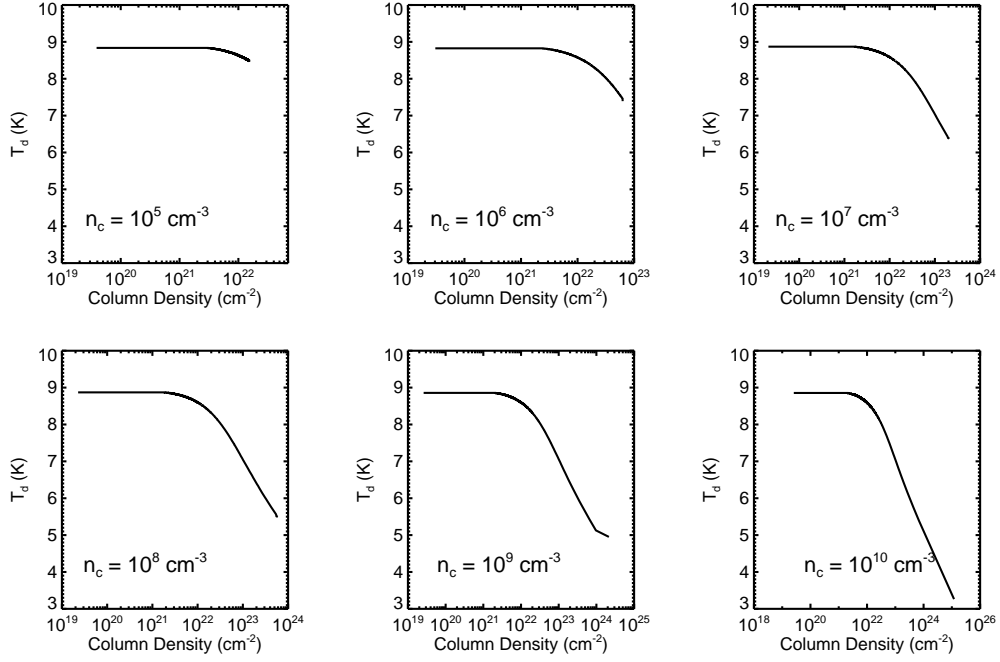}
\caption{\label{fig_be_empirical}$\bar{T_d}$, the weighted mean dust 
temperature along each line of sight, versus the column density along that 
same line of sight.  The six panels show the relationships for BE spheres 
ranging from $10^5$~cm$^{-3}$ to $10^{10}$~cm$^{-3}$, with each panel 
labeled in the lower left corner.}
\end{figure*}

Finally, we use these Bonnor-Ebert spheres to develop empirical 
relationships between column density and dust temperature, and use these 
relationships to assign dust temperatures to each pixel of the column 
density snapshots of the simulations (as described in \S \ref{sec_synthetic}).  
For each of the six BE sphere models considered here, in each pixel of the 
column density maps shown in Figure \ref{fig_sim_n_be} we calculate the 
mean dust temperature along the line-of-sight, weighted by the mass at each 
temperature.  In practice this weighted mean dust temperature, $\bar{T_d}$, is 
calculated as

\begin{equation}\label{eq_dust_temp}
\bar{T_d} = \frac{\int T(s) \, n(s) \, ds}{\int n(s) \, ds} \qquad ,
\end{equation}
where $s$ is the distance along the line of sight.  Figure 
\ref{fig_be_empirical} shows the resulting relationships between column 
density and dust temperature for each of the six BE sphere models.  
When assigning dust temperatures to each pixel in the column density images 
from the simulations, we select the BE sphere model with the closest maximum 
column density to that found in each timestep of the simulation and interpolate 
using the relationships shown in Figure \ref{fig_be_empirical} to assign a 
dust temperature to each value of the column density.

While dust temperatures below 10 K have been observed in starless cores 
\citep[e.g.,][]{evans2001:starless,bergin2006:b68}, these empirical 
relationships between column density and dust temperature are derived for BE 
spheres and may not be generally applicable to the simulations considered 
here.  
However, our results are only weakly sensitive to the exact 
temperature model adopted.  If, for example, we used the same model except 
with the addition of a 7 K temperature floor, the minimum central density at 
which the C04 simulation is detected decreases by a factor of three, from 
$n_c~=~8.9~\times~10^{7}$~cm$^{-3}$ to $n_c~=~3.2~\times~10^{7}$~cm$^{-3}$ 
(when viewed perpendicular to the magnetic field axis).  As a result, the 
total number of expected detections, averaged over all possible orientations 
of the magnetic field with respect to the line of sight, increases from at 
least two to at least three.

\clearpage

\input{tab1.tex}

\input{tab2.tex}

\input{tab3.tex}

\clearpage
\input{tab4.tex}
\clearpage
\input{tab5.tex}
\input{tab6.tex}

\end{document}

%% file: tab1.tex
\begin{deluxetable*}{lccccccc}
\tabletypesize{\scriptsize}
\tablewidth{0pt}
\tablecaption{\label{tab_observations}ALMA 106 GHz Continuum Observations}  
\tablehead{
               &                   &                   & \colhead{Synthesized}  & \colhead{Synthesized} &                       &                                       &                         \\
               & \colhead{R.A.}    & \colhead{Decl.}   & \colhead{Beam Size}    & \colhead{Beam P.A.} & \colhead{1$\sigma$ rms} & \colhead{$\sigma_M$\tablenotemark{a}} &  \colhead{Evolutionary} \\
\colhead{Source} & \colhead{(J2000)} & \colhead{(J2000)} & \colhead{(arcseconds)} & \colhead{(degrees)} & \colhead{(mJy beam$^{-1}$)} & \colhead{(\msun\ beam$^{-1}$)} & \colhead{Status\tablenotemark{b}}
}
\startdata
CHAI-01        & 11:08:03.22       & $-$77:39:17.9     & $2.72 \times 1.89$     & $-$40.2 & 0.12 & 0.0021   & D \\
CHAI-02        & 11:08:38.96       & $-$77:43:52.9     & $2.75 \times 1.92$     & $-$40.0 & 0.12 & 0.0021   & P \\
CHAI-03        & 11:10:00.13       & $-$76:34:59.1     & $2.67 \times 1.89$     & $-$42.4 & 0.11 & 0.0019   & D \\
CHAI-04        & 11:06:31.94       & $-$77:23:38.9     & $2.62 \times 1.82$     & $-$46.6 & 0.08 & 0.0014   & P \\
CHAI-05        & 11:06:46.44       & $-$77:22:32.2     & $2.71 \times 1.89$     & $-$40.7 & 0.12 & 0.0021   & P \\
CHAI-06        & 11:06:15.51       & $-$77:24:04.9     & $2.61 \times 1.82$     & $-$44.8 & 0.08 & 0.0014   & S \\
CHAI-07        & 10:58:16.92       & $-$77:17:18.3     & $2.74 \times 1.97$     & $-$39.2 & 0.12 & 0.0021   & D \\
CHAI-08        & 11:07:01.85       & $-$77:23:00.8     & $2.61 \times 1.82$     & $-$45.6 & 0.08 & 0.0014   & S \\
CHAI-09        & 11:04:16.41       & $-$77:47:32.1     & $2.63 \times 1.82$     & $-$44.8 & 0.08 & 0.0014   & S \\
CHAI-10        & 11:02:24.98       & $-$77:33:36.3     & $2.70 \times 1.89$     & $-$38.7 & 0.12 & 0.0021   & D \\
CHAI-11        & 11:08:50.40       & $-$77:43:46.7     & $2.63 \times 1.82$     & $-$45.0 & 0.09 & 0.0016   & S \\
CHAI-12        & 11:08:15.28       & $-$77:33:51.9     & $2.72 \times 1.89$     & $-$40.6 & 0.11 & 0.0019   & D \\
CHAI-13        & 11:08:00.07       & $-$77:38:43.8     & $2.60 \times 1.83$     & $-$41.5 & 0.10 & 0.0018   & D \\
CHAI-14        & 11:09:45.85       & $-$76:34:49.7     & $2.59 \times 1.82$     & $-$47.9 & 0.08 & 0.0014   & D \\
CHAI-15        & 11:06:59.93       & $-$77:22:00.6     & $2.61 \times 1.82$     & $-$45.9 & 0.08 & 0.0014   & S \\
CHAI-16        & 10:56:42.95       & $-$77:04:05.1     & $2.58 \times 1.82$     & $-$43.5 & 0.08 & 0.0014   & S \\
CHAI-17        & 10:56:30.56       & $-$77:11:39.8     & $2.65 \times 1.89$     & $-$38.8 & 0.11 & 0.0019   & D \\
CHAI-18        & 11:09:47.42       & $-$77:26:31.5     & $2.73 \times 1.89$     & $-$42.4 & 0.11 & 0.0019   & D \\
CHAI-20        & 11:07:21.34       & $-$77:22:12.1     & $2.72 \times 1.89$     & $-$41.4 & 0.11 & 0.0019   & D \\
CHAI-21        & 11:06:06.48       & $-$77:25:06.7     & $2.61 \times 1.82$     & $-$44.6 & 0.08 & 0.0014   & S \\
CHAI-22        & 11:04:23.35       & $-$77:18:07.8     & $2.69 \times 1.89$     & $-$39.2 & 0.12 & 0.0021   & P \\
CHAI-23        & 11:02:10.85       & $-$77:42:31.8     & $2.59 \times 1.82$     & $-$40.4 & 0.09 & 0.0016   & S \\
CHAI-24        & 11:09:26.24       & $-$76:33:28.3     & $2.58 \times 1.82$     & $-$47.1 & 0.08 & 0.0014   & P \\
CHAI-25        & 11:03:36.59       & $-$77:59:19.3     & $2.64 \times 1.82$     & $-$44.1 & 0.08 & 0.0014   & S \\
CHAI-26        & 11:07:17.13       & $-$77:23:20.2     & $2.60 \times 1.82$     & $-$44.5 & 0.08 & 0.0014   & S \\
CHAI-28        & 11:07:43.89       & $-$77:39:41.6     & $2.73 \times 1.89$     & $-$40.9 & 0.12 & 0.0021   & D \\
CHAI-29        & 11:10:01.05       & $-$76:36:32.9     & $2.57 \times 1.82$     & $-$46.3 & 0.08 & 0.0014   & S \\
CHAI-30        & 11:06:21.67       & $-$77:43:58.4     & $2.51 \times 1.80$     & $-$13.9 & 0.12 & 0.0021   & S \\
CHAI-32        & 11:03:21.97       & $-$77:36:04.4     & $2.50 \times 1.80$     & $-$13.6 & 0.13 & 0.0023   & S \\
CHAI-33        & 11:04:49.58       & $-$77:45:53.5     & $2.51 \times 1.80$     & $-$13.7 & 0.13 & 0.0023   & S \\
CHAI-34        & 11:03:39.07       & $-$77:47:54.0     & $2.51 \times 1.80$     & $-$13.5 & 0.13 & 0.0023   & S \\
CHAI-35        & 11:11:27.78       & $-$77:15:45.2     & $2.48 \times 1.80$     & $-$17.0 & 0.13 & 0.0023   & S \\
CHAI-36        & 11:04:07.89       & $-$77:48:13.1     & $2.51 \times 1.80$     & $-$13.0 & 0.13 & 0.0023   & S \\
CHAI-37        & 11:04:01.79       & $-$77:58:50.4     & $2.52 \times 1.80$     & $-$12.1 & 0.13 & 0.0023   & S \\
CHAI-38        & 11:03:04.43       & $-$77:34:19.4     & $2.50 \times 1.80$     & $-$12.4 & 0.13 & 0.0023   & S \\
CHAI-39        & 11:07:17.30       & $-$77:38:42.3     & $2.50 \times 1.80$     & $-$14.4 & 0.13 & 0.0023   & S \\
CHAI-40        & 11:09:03.86       & $-$77:43:34.6     & $2.51 \times 1.80$     & $-$13.9 & 0.13 & 0.0023   & S \\
CHAI-41        & 11:02:10.46       & $-$77:37:22.9     & $2.50 \times 1.80$     & $-$11.0 & 0.13 & 0.0023   & S \\
CHAI-42        & 11:09:39.48       & $-$76:36:04.6     & $2.44 \times 1.80$     & $-$16.4 & 0.14 & 0.0025   & S \\
CHAI-43        & 11:07:47.07       & $-$77:35:46.8     & $2.50 \times 1.80$     & $-$12.7 & 0.13 & 0.0023   & S \\
CHAI-44        & 11:05:57.64       & $-$77:24:18.9     & $2.49 \times 1.80$     & $-$13.5 & 0.13 & 0.0023   & S \\
CHAI-45        & 11:00:43.98       & $-$77:30:38.5     & $2.49 \times 1.80$     & $-$11.4 & 0.13 & 0.0023   & S \\
CHAI-46        & 11:00:58.29       & $-$77:29:41.2     & $2.59 \times 1.89$     & $-$24.8 & 0.11 & 0.0019   & S \\
CHAI-47        & 11:07:32.72       & $-$77:36:23.0     & $2.61 \times 1.89$     & $-$27.8 & 0.11 & 0.0019   & S \\
CHAI-49        & 11:02:33.40       & $-$77:42:20.1     & $2.61 \times 1.89$     & $-$25.6 & 0.11 & 0.0019   & S \\
CHAI-50        & 10:50:56.54       & $-$77:04:29.1     & $2.56 \times 1.89$     & $-$22.0 & 0.11 & 0.0019   & S \\
CHAI-52        & 11:09:15.17       & $-$77:18:10.7     & $2.59 \times 1.89$     & $-$26.7 & 0.11 & 0.0019   & S \\
CHAI-53        & 11:06:23.17       & $-$77:41:42.7     & $2.61 \times 1.89$     & $-$26.7 & 0.11 & 0.0019   & S \\
CHAI-54        & 11:09:56.09       & $-$77:14:16.7     & $2.59 \times 1.89$     & $-$27.5 & 0.11 & 0.0019   & S \\
CHAI-55        & 11:02:10.12       & $-$77:39:29.2     & $2.61 \times 1.89$     & $-$26.7 & 0.11 & 0.0019   & S \\
CHAI-56        & 11:10:03.64       & $-$76:38:02.6     & $2.56 \times 1.90$     & $-$30.0 & 0.11 & 0.0019   & S \\
CHAI-57        & 11:06:03.63       & $-$77:34:52.7     & $2.60 \times 1.89$     & $-$25.7 & 0.11 & 0.0019   & S \\
CHAI-58        & 11:02:58.44       & $-$77:38:42.5     & $2.61 \times 1.89$     & $-$26.7 & 0.11 & 0.0019   & S \\
CHAI-59        & 11:07:15.99       & $-$77:24:35.6     & $2.60 \times 1.89$     & $-$28.0 & 0.11 & 0.0019   & S \\
CHAI-60        & 11:09:51.10       & $-$76:51:19.2     & $2.57 \times 1.90$     & $-$27.8 & 0.11 & 0.0019   & S \\
CHAI-61        & 11:08:59.13       & $-$77:45:08.6     & $2.62 \times 1.89$     & $-$27.9 & 0.10 & 0.0018   & S \\
CHAI-64        & 11:11:22.58       & $-$77:16:24.4     & $2.59 \times 1.89$     & $-$27.8 & 0.11 & 0.0019   & S \\
CHAI-65        & 11:09:00.36       & $-$77:40:12.6     & $2.56 \times 1.87$     & $-$174.9 & 0.11 & 0.0019   & S \\
CHAI-67        & 11:07:24.50       & $-$77:39:40.6     & $2.56 \times 1.87$     & $-$173.6 & 0.12 & 0.0021   & S \\
CHAI-68        & 11:02:33.46       & $-$78:01:38.2     & $2.59 \times 1.87$     & $-$172.8 & 0.11 & 0.0019   & S \\
CHAI-70        & 11:04:47.42       & $-$77:44:58.1     & $2.57 \times 1.87$     & $-$172.2 & 0.12 & 0.0021   & S \\
CHAI-71        & 11:09:07.16       & $-$77:23:55.4     & $2.55 \times 1.87$     & $-$175.2 & 0.12 & 0.0021   & S \\
CHAI-72        & 11:07:35.16       & $-$77:21:45.8     & $2.53 \times 1.88$     & $+$9.9   & 0.10 & 0.0018   & S \\
CHAI-73        & 11:08:03.98       & $-$77:20:53.1     & $2.54 \times 1.87$     & $-$173.3 & 0.12 & 0.0021   & S \\
CHAI-74        & 11:10:23.04       & $-$76:38:40.4     & $2.50 \times 1.87$     & $-$174.6 & 0.11 & 0.0019   & S \\
CHAI-77        & 11:11:09.52       & $-$76:42:00.2     & $2.70 \times 1.89$     & $-$43.8  & 0.12 & 0.0021   & D \\
CHAI-78        & 11:07:28.55       & $-$77:19:40.6     & $2.52 \times 1.88$     & $+$10.1  & 0.11 & 0.0019   & S \\
CHAI-79        & 11:06:18.59       & $-$77:44:40.1     & $2.57 \times 1.87$     & $-$173.4 & 0.12 & 0.0021   & S \\
CHAI-80        & 11:05:28.19       & $-$77:41:18.1     & $2.57 \times 1.87$     & $-$173.3 & 0.12 & 0.0021   & S \\
CHAI-81        & 10:56:49.93       & $-$77:03:02.8     & $2.53 \times 1.87$     & $+$10.4  & 0.12 & 0.0021   & S \\
CHAI-82        & 11:02:12.48       & $-$77:41:19.1     & $2.57 \times 1.88$     & $+$6.5   & 0.12 & 0.0021   & S \\
CHAI-83        & 11:03:00.75       & $-$77:49:18.2     & $2.57 \times 1.87$     & $-$173.6 & 0.12 & 0.0021   & S \\
CHAI-84        & 11:09:53.67       & $-$76:42:12.0     & $2.49 \times 1.87$     & $+$10.7  & 0.13 & 0.0023   & S 
\enddata
\tablenotetext{a}{1$\sigma$ mass sensitivity.  See text for details.}
\tablenotetext{b}{P -- protostellar core.  S -- starless core.  D -- disk, not considered a core (see text for details).  }
\end{deluxetable*}

%% file: tab2.tex
\begin{deluxetable*}{lccccccc}
\tabletypesize{\scriptsize}
\tablewidth{0pt}
\tablecaption{\label{tab_continuum}Observed Properties of ALMA 106 GHz Continuum Detections}  
\tablehead{
 &                &                 & \colhead{Deconvolved} & \colhead{Deconvolved} & \colhead{Deconvolved} & \colhead{Peak} & \colhead{Integrated} \\
 & \colhead{R.A.} & \colhead{Decl.} & \colhead{Major Axis\tablenotemark{a}} & \colhead{Minor Axis\tablenotemark{a}} & \colhead{P.A.\tablenotemark{a}} & \colhead{Flux Density} & \colhead{Flux Density} \\
\colhead{Source} & \colhead{(J2000)} & \colhead{(J2000)} & \colhead{(arcsec)} & \colhead{(arcsec)} & \colhead{(degrees)} & \colhead{(mJy beam$^{-1}$)} & \colhead{(mJy beam$^{-1}$)}
}
\startdata
CHAI-01   & 11:08:03.21 & $-$77:39:17.4 & 1.42 (0.01) & 1.04 (0.01) &   7.1 (1.2)  & 73.9 (0.1) & 97.2 (0.2) \\
CHAI-02-A & 11:08:38.55 & $-$77:43:52.2 & 2.01 (0.02) & 0.88 (0.05) &   5.1 (1.1)  & 26.6 (0.2) & 38.5 (0.2) \\
CHAI-02-B & 11:08:43.12 & $-$77:43:50.6 & U           & U           &   U          &  2.0 (0.2) &  2.1 (0.2) \\
CHAI-03   & 11:09:59.98 & $-$76:34:58.0 & 0.51 (0.02) & 0.34 (0.03) &  32.1 (5.3)  & 64.9 (0.1) & 67.7 (0.1) \\
CHAI-04   & 11:06:33.13 & $-$77:23:33.8 & 3.30 (0.19) & 1.54 (0.36) & 145.3 (3.8)  &  1.7 (0.1) &  3.5 (0.3) \\
CHAI-05   & 11:06:46.43 & $-$77:22:32.9 & 0.56 (0.11) & 0.47 (0.14) &  21.0 (52.0) & 17.6 (0.2) & 18.6 (0.2) \\
CHAI-07   & 10:58:16.64 & $-$77:17:17.2 & 0.99 (0.08) & 0.71 (0.12) & 128.0 (17.0) &  9.7 (0.1) & 11.0 (0.1) \\
CHAI-08-A & 11:06:57.97 & $-$77:22:48.4 & U           & U           &   U          &  0.7 (0.1) &  0.7 (0.1) \\
CHAI-08-B & 11:07:09.61 & $-$77:23:04.9 & U           & U           &   U          &  1.3 (0.1) &  1.1 (0.1) \\
CHAI-10   & 11:02:24.79 & $-$77:33:35.6 & U           & U           &   U          & 13.0 (0.1) & 13.1 (0.1) \\
CHAI-12   & 11:08:15.37 & $-$77:33:53.4 & U           & U           &   U          &  6.5 (0.1) &  6.4 (0.1) \\
CHAI-13-A & 11:07:57.85 & $-$77:38:45.0 & U           & U           &   U          &  1.7 (0.1) &  1.7 (0.1) \\
CHAI-13-B & 11:08:02.85 & $-$77:38:42.6 & U           & U           &   U          &  7.6 (0.1) &  7.3 (0.1) \\
CHAI-14-A & 11:09:41.83 & $-$76:34:58.5 & 0.63 (1.96) & 0.56 (1.26) & 159.0 (65.0) &  3.4 (0.1) &  3.7 (0.1) \\
CHAI-14-B & 11:09:53.28 & $-$76:34:25.2 & U           & U           &   U          &  3.3 (0.3) &  3.1 (0.3) \\
CHAI-17   & 10:56:30.27 & $-$77:11:39.3 & 1.35 (0.08) & 0.87 (0.13) & 164.3 (9.5)  &  6.2 (0.1) &  7.7 (0.1) \\
CHAI-18   & 11:09:47.29 & $-$77:26:29.2 & 0.94 (1.79) & 0.41 (1.48) &  15.0 (57.0) &  7.0 (0.1) &  7.8 (0.1) \\
CHAI-20   & 11:07:21.39 & $-$77:22:11.7 & 0.72 (1.99) & 0.40 (1.48) &  26.0 (67.0) &  7.0 (0.1) &  7.5 (0.1) \\
CHAI-22-A & 11:04:22.64 & $-$77:18:08.2 & U           & U           &   U          &  4.0 (0.1) &  3.9 (0.1) \\
CHAI-22-B & 11:04:23.22 & $-$77:18:07.0 & U           & U           &   U          &  1.5 (0.1) &  2.0 (0.1) \\
CHAI-24   & 11:09:28.40 & $-$76:33:28.2 & 1.61 (0.96) & 0.84 (0.98) & 158.0 (36.0) &  1.1 (0.1) &  1.5 (0.1) \\
CHAI-28   & 11:07:43.53 & $-$77:39:41.1 & U           & U           &   U          &  6.1 (0.1) &  6.3 (0.1) \\
CHAI-39   & 11:07:20.52 & $-$77:38:06.6 & 5.12 (0.37) & 0.57 (0.51) & 175.5 (4.6)  &  2.7 (0.5) &  6.5 (1.3) \\
CHAI-77   & 11:11:10.75 & $-$76:41:57.2 & U           & U           &   U          &  3.0 (0.1) &  3.2 (0.1) 
%CHAI-08-B & 11:06:59.71 & $-$77:22:50.4 & U   & U   &   U & 0.3  & 0.7  \\
%CHAI-57   & 11:05:55.86 & $-$77:34:42.4 & U           & U           &   U          &  0.7 (0.4) &  1.5 (0.9) \\
\enddata
\tablenotetext{}{Statistical uncertainties are listed in parentheses.}
\tablenotetext{a}{U - Unresolved in one or both dimensions and thus unable to be deconvolved with the beam.}
\end{deluxetable*}

%% file: tab3.tex
\begin{deluxetable}{lccc}
\tabletypesize{\scriptsize}
\tablewidth{0pt}
\tablecaption{\label{tab_continuum_physical}Physical Properties of ALMA 106 GHz Continuum Detections}  
\tablehead{
                 & \colhead{Effective} &                   & \\
                 & \colhead{Radius}    & \colhead{Mass}    & \colhead{$n$} \\
\colhead{Source} & \colhead{(AU)}      & \colhead{(\msun)} & \colhead{(cm$^{-3}$)}
}
\startdata
CHAI-01    &  91 (1)  & 1.7 (0.0035)     & 8.1 (0.3) $\times 10^{10}$ \\
CHAI-02-A  & 100 (3)  & 0.68 (0.0035)    & 2.4 (0.2) $\times 10^{10}$ \\
CHAI-02-B  & $<$172   & 0.037 (0.0035)   & $>$2.6 $\times 10^{8}$ \\
CHAI-03    &  31 (3)  & 1.2 (0.0018)     & 1.4 (0.4) $\times 10^{12}$ \\
CHAI-04    & 169 (20) & 0.062 (0.005)    & 4.6 (1.7) $\times 10^{8}$ \\
CHAI-05    &  38 (7)  & 0.33 (0.0035)    & 2.2 (1.2) $\times 10^{11}$ \\
CHAI-07    &  63 (6)  & 0.19 (0.0018)    & 2.7 (0.8) $\times 10^{10}$ \\
CHAI-08-A  &  $<$164  & 0.012 (0.0018)   & $>$9.8 $\times 10^{7}$ \\
CHAI-08-B  &  $<$164  & 0.019 (0.0018)   & $>$1.5 $\times 10^{8}$ \\
CHAI-10    &  $<$169  & 0.23 (0.0018)    & $>$1.7 $\times 10^{9}$ \\
CHAI-12    &  $<$171  & 0.11 (0.0018)    & $>$7.9 $\times 10^{8}$ \\
CHAI-13-A  &  $<$164  & 0.030 (0.0018)   & $>$2.4 $\times 10^{8}$ \\
CHAI-13-B  &  $<$164  & 0.13 (0.0018)    & $>$1.1 $\times 10^{9}$ \\
CHAI-14-A  &  45 (86) & 0.066 (0.0018)   & 2.6 (15) $\times 10^{10}$ \\
CHAI-14-B  &  $<$163  & 0.055 (0.0053)   & $>$4.6 $\times 10^{8}$ \\
CHAI-17    &  82 (7)  & 0.14 (0.0018)    & 9.1 (2.3) $\times 10^{9}$ \\
CHAI-18    &  47 (95) & 0.14 (0.0018)    & 4.8 (29) $\times 10^{10}$ \\
CHAI-20    &  40 (93) & 0.13 (0.0018)    & 7.3 (51) $\times 10^{10}$ \\
CHAI-22-A  &  $<$169  & 0.069 (0.0018)   & $>$5.1 $\times 10^{8}$ \\
CHAI-22-B  &  $<$169  & 0.035 (0.0018)   & $>$2.6 $\times 10^{8}$ \\
CHAI-24    &  87 (57) & 0.027 (0.0018)   & 1.5 (2.9) $\times 10^{9}$ \\
CHAI-28    &  $<$170  & 0.11 (0.0018)    & $>$8.0 $\times 10^{8}$ \\
CHAI-39    & 128 (58) & 0.12 (0.023)     & 2.1 (2.8) $\times 10^{9}$ \\
CHAI-77    &  $<$169  & 0.057 (0.0018)   & $>$4.2 $\times 10^{8}$ 
\enddata
\tablenotetext{}{Statistical uncertainties are listed in parentheses.}
\end{deluxetable}

%% file: tab4.tex
\begin{landscape}
\begin{deluxetable}{lcccccccc}
\tabletypesize{\scriptsize}
\tablewidth{0pt}
\tablecaption{\label{tab_associations}Associations with Known Sources}  
\tablehead{
                 & \multicolumn{2}{c}{\textbf{\underline{LABOCA Sources \citep{belloche2011:chami}}}}  & \multicolumn{3}{c}{\textbf{\underline{\herschel\ YSOs \citep{winston2012:chami}}}} & \multicolumn{3}{c}{\textbf{\underline{\spitzer\ Protostars \citep{dunham2013:luminosities}}}} \\
\colhead{}       & \colhead{d\tablenotemark{a}} & \colhead{}                             & \colhead{}               & \colhead{d\tablenotemark{a}} & \colhead{}                       & \colhead{Protostar}   & \colhead{d\tablenotemark{a}} & \colhead{} \\
\colhead{Source} & \colhead{(\arcsec)}  & \colhead{Associated?\tablenotemark{b}} & \colhead{YSO (2MJ $+$)\tablenotemark{c}} & \colhead{(\arcsec)}  & \colhead{Class\tablenotemark{d}} & \colhead{(SSTgb $+$)\tablenotemark{e}} & \colhead{(\arcsec)}  & \colhead{Explanation\tablenotemark{f}}
}
\startdata
CHAI-01   & 0.5  & Y & 11080329$-$7739174 & 0.3     & II      & None              & \nodata & Saturated    \\
CHAI-02-A & 1.5  & Y & 11083896$-$7743513 & 1.6     & I       & None              & \nodata & Saturated    \\
CHAI-02-B & 13.5 & N & None               & \nodata & \nodata & None              & \nodata & Not resolved \\
CHAI-03   & 1.2  & Y & 11100010$-$7634578 & 0.5     & II      & None              & \nodata & Saturated    \\
CHAI-04   & 6.4  & Y & 11063460$-$7723340 & 4.8     & 0       & None              & \nodata & Not detected \\
CHAI-05   & 0.9  & Y & 11064658$-$7722325 & 0.8     & I       & J1106464$-$772232 & 0.2     & \nodata      \\
CHAI-07   & 1.4  & Y & 10581677$-$7717170 & 1.4     & II      & None              & \nodata & Coverage     \\
CHAI-08-A & 17.8 & N & None               & \nodata & \nodata & J1106580$-$772248 & 0.5     & \nodata      \\
CHAI-08-B & 26.0 & N & 11070919$-$7723049 & 1.4     & Flat    & None              & \nodata & No core      \\
CHAI-10   & 0.9  & Y & 11022491$-$7733357 & 0.4     & TD      & None              & \nodata & Coverage     \\
CHAI-12   & 1.5  & Y & 11081509$-$7733531 & 1.0     & Flat    & None              & \nodata & Saturated    \\
CHAI-13-A & 7.2  & Y & 11075792$-$7738449 & 0.3     & Flat    & None              & \nodata & No core      \\
CHAI-13-B & 9.0  & Y & 11080297$-$7738425 & 0.4     & II      & J1108029$-$773842 & 0.5     & \nodata      \\
CHAI-14-A & 16.5 & N & 11094192$-$7634584 & 0.3     & II      & None              & \nodata & No core      \\
CHAI-14-B & 35.6\tablenotemark{g} & N & 11095340$-$7634255 & 0.5     & II      & None              & \nodata & No core      \\
CHAI-17   & 1.1  & Y & 10563044$-$7711393 & 0.6     & II      & None              & \nodata & Coverage     \\
CHAI-18   & 2.3  & Y & 11094742$-$7726290 & 0.5     & II      & J1109472$-$772629 & 0.3     & \nodata      \\
CHAI-20   & 0.4  & Y & 11072142$-$7722117 & 0.1     & Flat    & J1107213$-$772211 & 0.1     & \nodata      \\
CHAI-22-A & 2.4  & Y & 11042275$-$7718080 & 0.4     & I       & J1104227$-$771808 & 0.3     & \nodata      \\
CHAI-22-B & 2.2  & Y & 11042275$-$7718080 & 1.8     & I       & None              & \nodata & Not resolved \\
CHAI-24   & 7.5  & Y & 11092855$-$7633281 & 0.5     & I       & J1109285$-$763328 & 0.4     & \nodata      \\
CHAI-28   & 1.3  & Y & 11074366$-$7739411 & 0.4     & II      & J1107435$-$773941 & 0.4     & \nodata      \\
CHAI-39   & 37.2 & N & 11072074$-$7738073 & 1.0     & II      & None              & \nodata & No core      \\
CHAI-77   & 5.2  & Y & 11111083$-$7641574 & 0.3     & II      & J1111107$-$764157 & 0.3     & \nodata
%CHAI-08-B & 12.6 & N & None               & \nodata & \nodata & None              & \nodata & No source    \\
%CHAI-57   & 27.1 & N & None               & \nodata & \nodata & None              & \nodata & No source    \\
\enddata
\tablenotetext{a}{Projected separation between ALMA continuum detection and LABOCA source, {\it Herschel} YSO, or {\it Spitzer} protostar.}
\tablenotetext{b}{Yes (Y) or no (N) depending on whether the projected separation is within the half-power radius of the LABOCA beam (10.6\arcsec); see text for details.}
\tablenotetext{c}{The prefix 2MJ should be appended to each tabulated source name.}
\tablenotetext{d}{Evolutionary class as listed by \citet{winston2012:chami}.  0 - Class 0.  I - Class I.  Flat - Flat-spectrum object.  II - Class II.  TD - transition disk.}
\tablenotetext{e}{The prefix SSTgb should be appended to each tabulated source name.}
\tablenotetext{f}{Explanation for why there is no associated \citet{dunham2013:luminosities} protostar, if applicable (see text for more details).  Saturated - saturated at one or more wavelengths. Not resolved - not resolved into separate sources at one or more wavelengths.  Not detected - not detected at one or more wavelengths.  Coverage - not covered at one or more wavelengths.  No core - not associated with a (sub)millimeter continuum core.}
\tablenotetext{g}{CHAI-14-B is located 2.9\arcsec\ away from CHAI-62 in \citet{belloche2011:chami}, one of the three sources tentatively associated with a young stellar object and omitted from our survey for reasons discussed in \S \ref{sec_observations}.  It is located close enough to CHAI-14 to be covered in the same pointing.}
\end{deluxetable}
\clearpage
\end{landscape}

%% file: tab5.tex
\begin{deluxetable*}{lccccc}
\tabletypesize{\scriptsize}
\tablewidth{0pt}
\tablecaption{\label{tab_fhsc}Peak Flux Densities of Candidate First Hydrostatic Cores in Perseus}  
\tablehead{
                 &                          &                     &                     &                     &                                \\
                 & \colhead{Observed}       &                     &                     &                     &                                \\
                 & \colhead{Peak Intensity} &                     & \multicolumn{3}{c}{\underline{Predicted Peak Intensity at 106 GHz (\mjybeam)\tablenotemark{b}}} \\
\colhead{Object} & \colhead{(\mjybeam)}     & \colhead{Reference\tablenotemark{a}} & \colhead{$\beta=2$} & \colhead{$\beta=1$} & \colhead{$\beta=0$} 
}
\startdata
L1448-IRS2E    & 6.0 (225 GHz)   & 1 & 0.7 & 1.5 & 3.1 \\
Per-Bolo 45    & 2.4 (102 GHz)   & 2 & 6.6 & 6.3 & 6.1 \\
Per-Bolo 58    & 6.9 (225 GHz)   & 3 & 0.8 & 1.7 & 3.6 \\
L1451-mm       & 2.0 (90 GHz)    & 4 & 9.0 & 7.7 & 6.5 \\
L1451-mm       & 32.8 (225 GHz)  & 4 & 3.8 & 8.1 & 17.1 
\enddata
\tablenotetext{a}{References for observed peak intensities:  (1) \citet{chen2010:fhsc}; (2) \citet{schnee2010:starless}; (3) \citet{dunham2011:fhsc}; (4) \citet{pineda2011:fhsc}.}
\tablenotetext{b}{Scaled to the distance of Chamaeleon~I.  See text for details on how these predictions are calculated.}
\end{deluxetable*}

%% file: tab6.tex
%Need to add initial density?
\begin{deluxetable}{lcccccccc}
\tablecolumns{6}
\tablecaption{Simulation Properties \label{simprop}}
\tablehead{ 
\colhead{}                       & \colhead{$M$}     & \colhead{$R_c$} & \colhead{$n$} & \colhead{$B$}      & \colhead{$\sigma$} & \colhead{$\Delta x_{\rm max}$} \\
\colhead{Simulation\tablenotemark{a}} & \colhead{(\msun)} & \colhead{(AU)}  & \colhead{(cm$^{-3}$)} & \colhead{($\mu$G)} & \colhead{(\kms)}   & \colhead{(AU)}
}
\startdata
C4 & 4.0 & $1.3 \times 10^{4}$ & $6.0 \times 10^{4}$ & 20.6 & 0.26 & 3.3 \\
C04 & 0.4 & $4.4 \times 10^{3}$ & $1.6 \times 10^{5}$ & 18.54 & 0.16 &1.1  %rho = 6.44d-19
\enddata
\tablenotetext{a}{
The model name, total gas mass, initial core radius, initial (uniform) number density, magnetic field strength, initial rms velocity dispersion, and simulation resolution on the maximum level. The magnetic field strength corresponds to a mass-to-flux ratio five times the critical value. }
\end{deluxetable}